# Wake Fields and Impedance

L. Palumbo ◊*), V.G. Vaccaro @§) and M. Zobov ◊)

◊) INFN-LNF Frascati
§) INFN, Sezione di Napoli
∗) Università di Roma "La Sapienza"
@) Università di Napoli "Federico II"

**Abstract**

Knowledge of the electromagnetic interaction between a beam and the surrounding vacuum chamber is necessary in order to optimize the accelerator performance in terms of stored current. Many instability phenomena may occur in the machine because of the fields produced by the beam and acting back on itself as in a feedback device. Basically, these fields produce an extra voltage and energy gain, affecting the longitudinal dynamics, and a transverse momentum kick which deflects the beam. In this paper we describe the main features of this interaction with typical machine components.

# 1. INTRODUCTION

The so called "collective effects" are responsible of many phenomena which limit the performance of an accelerator in terms of beam quality and stored current. The beam traveling inside a complicated vacuum chamber, induces electromagnetic fields which may affect the dynamics of the beam itself. An accelerator can be seen therefore as a feedback device, where any longitudinal or transverse perturbation appearing in the beam distribution may be amplified (or damped) by the e.m. forces generated by the perturbation itself.

The e.m. fields induced by the beam are referred to as *wake fields* due to the fact that they are left mainly behind the traveling charge. In the limit case of a charge moving at the light velocity, $\beta=1$, the fields can only stay behind the charge because of the causality principle.

The study of the longitudinal and transverse beam dynamics requires the knowledge of the forces acting on the beam or, alternatively, the change in momentum caused by these e.m. forces. The *longitudinal wake potential* (*volts*) is the voltage gain of a unit trailing charge due to the fields created by a leading charge. The *transverse wake potential* (*volts*) is the transverse momentum kick experienced by the beam because of the deflecting fields. They are sometimes confused with the *wake function*s, defined as the wake potentials per unit charge (*volt /coulomb*) defining, therefore, a Green's function for the problem.

When we study the beam dynamics in the time domain, as usually done for linear accelerators, it is convenient to make use of the wake functions or potentials. On the other side the frequency domain analysis is usually adopted for circular accelerators due to the intrinsic periodicity. There we need to compute the frequency Fourier transform of the wake function, which having Ohms units, is called *coupling impedance*.

In this paper we describe the main features of the electromagnetic fields induced in the most typical components installed on the beam pipe of an accelerator. In some examples we make use of numerical codes, reliable tools for the estimate of wake potentials and impedances, particularly useful in the design of the machine components. On this subject we address the readers to Ref. [1] where an exhaustive review on the available computer codes is presented. Methods and techniques for measurements of wake potentials and impedance are described in Ref. [2].

# 2. LONGITUDINAL WAKE FUNCTION AND LOSS FACTOR

## 2.1 Longitudinal wake function and loss factor of a point charge

Let us consider a charge $q_1$ traveling with constant velocity $v = \beta c$ on trajectories parallel to the axis of a vacuum chamber. Let $z_1$ be the longitudinal position and $\boldsymbol{r}_1$ the transverse vector positions (Fig. 1).

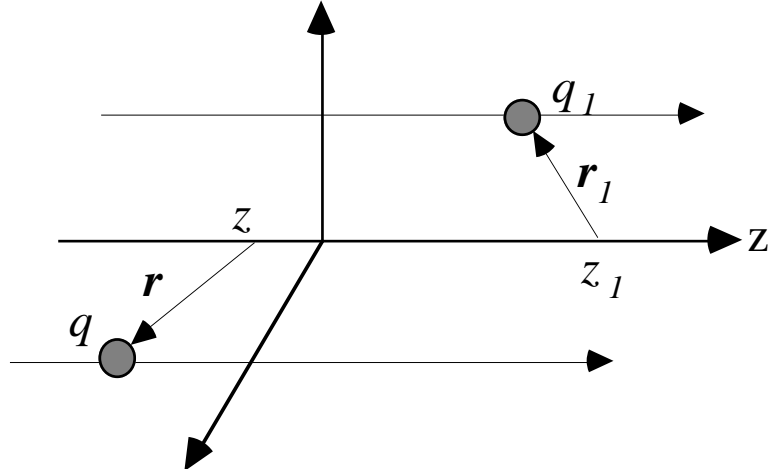

Fig. 1 - Relevant coordinates system

The electromagnetic fields **E** and **B** produced by the charge $q_1$ in the structure can be derived by solving the Maxwell equations satisfying proper boundary conditions. The Lorentz force acting on a charge $q$ at a given position $r,z$:

$$\boldsymbol{F}(r,z,r_1,z_1\ ;\ t) = q[\boldsymbol{E}(r,z,r_1,z_1\ ;\ t) + \boldsymbol{v} \times \boldsymbol{B}(r,z,r_1,z_1\ ;\ t)] \tag{1}$$

has in general field components along and perpendicular to the trajectory. These e.m. fields affect the dynamics of the charge itself and on any trailing charge as well. Calling $\tau$ the time delay of the trailing charge with respect to the leading one, at any instant "t" the leading and trailing charges have longitudinal coordinates $z_1(t) = vt$ and $z(t) = v(t-\tau)$ respectively.

The energy lost by the charge $q_1$ is computed as the work done by the longitudinal e.m. force along the structure:

$$U_{11}(\boldsymbol{r}_1) = -\int_{-\infty}^{\infty} \boldsymbol{F}(\boldsymbol{r}_1,z_1,\boldsymbol{r}_1,z_1\ ;\ t)\cdot dz \quad;\quad t = \frac{z_1}{v} \tag{2}$$

The quantity $U_{11}$ accounts for the energy loss in the resistive walls and in the diffracted fields radiated caused by the discontinuities of the vacuum pipe. For a point charge, apart of particular cases, it is generally positive (energy loss).

The trailing charge also changes its energy under the effect of the fields produced by the leading one:

$$U_{21}(\boldsymbol{r},\boldsymbol{r}_1;\tau) = -\int_{-\infty}^{\infty} \boldsymbol{F}(\boldsymbol{r},z,\boldsymbol{r}_1,z_1\ ;\ t)\cdot dz \quad;\quad t = \frac{z_1}{v} + \tau \tag{3}$$

where the force is calculated on the charge q, on the same path but with a time delay $\tau$.

The quantity $U_{21}$ depending on the time delay $\tau$ can be positive (energy loss) or negative (energy gain). As long as we consider charges moving on trajectories parallel to the z-axis, the magnetic field cannot change the particles energy, the product $\mathbf{v} \times \mathbf{B} \cdot d\mathbf{z} = \mathbf{0}$ being identically zero. Accordingly, the energy gain of Eqs. (2) and (3) is computed considering the longitudinal component of the electric field only.

In the above definitions we have considered the integration over an infinite path. Of course infinite structures do not exist in practice, neither in linac nor in accelerator rings. In real machine components we may have fields confined in a limited region (for examples resonant fields below the beam pipe cut off), or propagating into the vacuum chamber. Extension of the integration path over an infinite pipe is certainly allowed in the former case. In the latter, definition (3) gives an estimate of the energy gain, which is a good approximation as long as the field wavelength is short compared to the device length.

A real vacuum chamber is formed by a smooth beam pipe with regular cross section (circular, rectangular or elliptic) and by various devices such as the RF cavity, the kickers, the diagnostic components etc. The exact solution of Maxwell equation for the whole structure is impossible to obtain, even with the most sophisticated computer codes. Usually, one analyses a component at a time and sum up the various effects. This procedure may lead to inexact estimates at high frequency where interference effects are not negligible.

It has to be underlined that in Eqs. (2) and (3) we assume the charge velocity unchanged during the motion. One can imagine that an external force keeps constant the charge velocity doing the work computed in Eqs. (2,3). In absence of the external force, this work corresponds to the energy loss (or gain) of the charges, provided that the velocity of the charge does not change significantly. In practice Eqs. (2,3) may be used when the relative change of energy is very small, such not to produce an appreciable variation of the relativistic factor $\beta$. This is the case, for instance, of ultra-relativistic charges. Otherwise, one has to introduce the equations of the dynamics combined to Maxwell equations.

We define *loss factor* $k$ the energy lost by $q_1$ per unit charge squared:

$$k(\mathbf{r}_1) = \frac{U_{11}(\mathbf{r}_1)}{q_1^2} \qquad [\text{V/C}] \qquad (4)$$

and *longitudinal wake function* $w_z(\mathbf{r}_1, \mathbf{r}_2; \tau)$ the energy lost by the trailing charge $q$ per unit of both charges $q_1$ and $q$ [3,4,5]:

$$w_z(\mathbf{r}, \mathbf{r}_1; \tau) = \frac{U_{21}(\mathbf{r}, \mathbf{r}_1; \tau)}{q_1 q} \qquad [\text{V/C}] \qquad (5)$$

The explicit dependence on $\beta$ although omitted, should be borne in mind. We note that both the wake function and the loss factor have the same units *volt/coulomb*. Sometimes in the literature one finds that the quantity $w(\tau)$ is unproperly called wake potential; the wake function is numerically equal to the potential seen by the charge only when one considers unity charges.

In some cases, such as infinite beam pipe with perfectly conducting or resistive walls, the e.m. force is constant along the integration path; it is therefore useful to introduce the wake function per unit length, *volt/(coulomb meter)*, given by:

$$\frac{dw_z(\mathbf{r},\mathbf{r}_1,\tau)}{dz} = -\frac{1}{q_1 q} F_z(\mathbf{r},z,\mathbf{r}_1,z_1; t) \quad ; \quad z = z_1 - v\tau \qquad [\text{V/Cm}] \qquad (6)$$

which, apart of the sign, is in practice the longitudinal force per unit charge acting on *q*. In some other cases where we deal with periodic structures, we rather calculate a wake force per unit period length.

It is worth noting that in the most cases of interest we deal with structures having particular symmetric shapes: rectangular, elliptic, circular. Moreover, it is generally verified that during the machine operation the beam can only slightly be displaced from the axis. Accordingly the above quantities can be expanded around the axis keeping only few relevant terms. This multipolar expansion assumes a particular form in case of cylindrical symmetry and ultra relativistic charges, as it will be shown in Sec. 2.6. The dominant term produced by a charge on the axis is called *monopole wake* (Fig. 2).

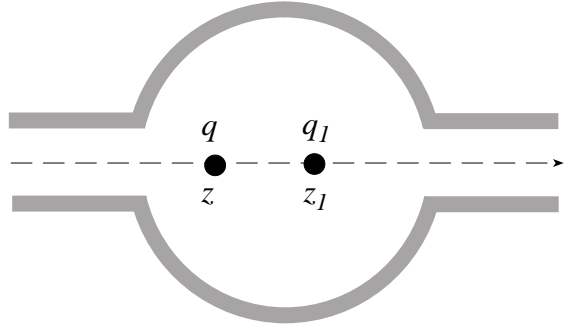

Fig. 2 - Leading and trailing charges on the axis of a cavity with cylindrical symmetry

## 2.2 Beam loading theorem for a point charge

From the above definitions we easily derive that, when the charges travel on the same trajectory, the loss factor is given by the wake function in the limit of zero distance between $q_1$ and $q$. Omitting the radial dependence, one obtains : $k = w_z(0)$. This is generally true as long as $\beta < 1$, however, in the relevant case $\beta = 1$ it has been proved that [3]:

$$k = \frac{w_z(\tau \to 0^+)}{2} \qquad (7)$$

This property, referred to as fundamental theorem of the beam loading [3], is a consequence of the causality principle. In fact, due to the finite propagation velocity of the induced fields and to the motion of the source charge, the wake function is not symmetric with respect to the leading charge (Fig. 3a). In the limit case of a charge with light velocity it exists only in the region $\tau > 0$ (Fig. 3b), showing a discontinuity at the origin.

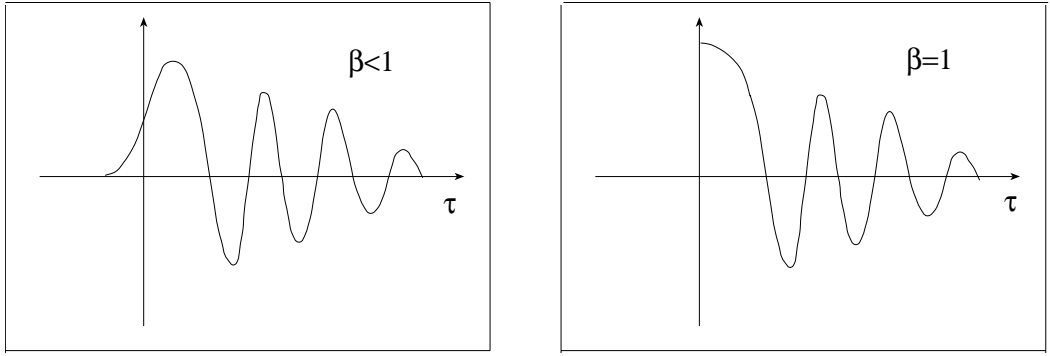

Fig. 3 - Example of wake functions for a) $\beta < 1$, and b) $\beta = 1$.

To prove the theorem, let us consider the wake function produced by a point charge as the sum of an even and odd function of $\tau$ (Fig. 4):

$$w_z(\tau) = w_z^e(\tau) + w_z^o(\tau) \tag{8}$$

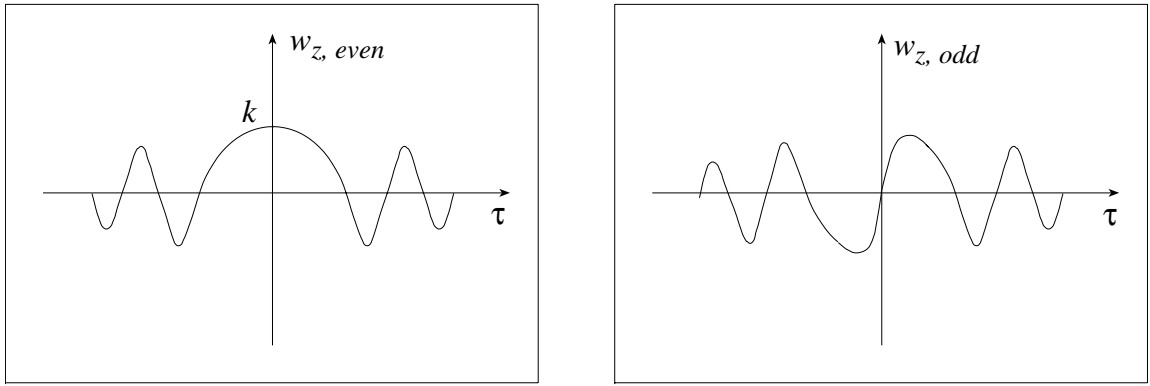

Fig. 4 - Even and odd part of the wake of Fig. 3a ($\beta < 1$)

It is apparent that only $w_z^e(\tau)$ may change the energy of the point charge, $w_z^o(\tau)$ being zero at $\tau=0$. Therefore we can say that the loss factor of a point charge is given by:

$$k = w_z^e(\tau = 0) \tag{9}$$

For $\beta = 1$, we have that $w_z(\tau) = 0$ for $\tau<0$, because of the causality principle. In this region the wake vanishes if:

$$w_z^e(\tau) = -w_z^o(\tau) \tag{10}$$

On the other side we have, for $\tau > 0$

$$w_z^e(\tau) = w_z^o(\tau) \tag{11}$$

$$w_z(\tau) = 2w_z^e(\tau) = 2w_z^o(\tau) \tag{12}$$

therefore from Eq.(9) we get:

$$k = w_z^e(\tau \to 0) = \frac{w_z(\tau \to 0^+)}{2} \tag{13}$$

We call the reader's attention on the fact that in general, as long as $\beta < 1$, i.e. in all the realistic cases, the wake is a continuous function of $\tau$. Therefore, it is more than reasonable to wonder about the meaning of Eq. (13) that applies only in the unrealistic case $\beta = 1$. It is easy to see that although the wake be a continuous function for any realistic value of $\beta$, its shape approaches more and more the discontinuous curve of Fig. 3 when $\beta \to 1$. In other words, one could not, in principle, exchange the limits $\tau \to 0$ and $\beta \to 1$.

2.2.1 Example: Point charge wake for a single resonating mode HOM.

As it will be shown in Sec. 7.4, a point charge $q_1$ passing through a resonant cavity excites all the resonating modes. In the limit case $\beta = 1$, each mode is schematized by the electric RLC parallel circuit driven by a point charge current $i_b(\tau) = q_1 \delta(\tau)$:

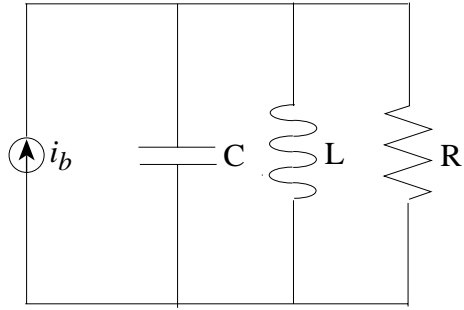

Fig. 5 - Scheme of a RLC parallel circuit driven by the current $i_b(\tau)$

At the time $\tau = 0^+$ we observe that the capacitor is charged with a voltage:

$$V(0^+) = \frac{q_1}{C} \equiv V_o \text{ and } \dot{V}(0^+) = \frac{V(0^+)}{RC} \tag{14}$$

For $t > 0$ the system will undergo free oscillations. In particular the voltage $V(\tau)$ will be a solution of differential equation of the circuit:

$$\ddot{V}(\tau) + 2\Gamma\dot{V}(\tau) + \omega_r^2 V(\tau) = 0 \tag{15}$$

where:

$$2\Gamma = \frac{1}{RC}, \quad \text{and} \quad \omega_r^2 = \frac{1}{LC} \tag{16}$$

Solving Eq.(15) with the initial conditions (14) and according to the definition (5), one gets:

$$w_z(\tau) = \frac{V(\tau)}{q_1} = \frac{e^{-\Gamma\tau}}{C}\left[\cos(\overline{\omega}_r\tau) - \frac{\Gamma}{\overline{\omega}_r}\sin(\overline{\omega}_r\tau)\right] H(\tau) \tag{17}$$

where $H(\tau)$ is the Heaveside function and

$$\overline{\omega}_r^2 = \omega_r^2 - \Gamma^2 \tag{18}$$

Using the merit factor of the circuit defined by:

$$Q = \frac{\omega_r}{2\Gamma} \tag{19}$$

$$\begin{aligned} w_o &= \frac{1}{C} = \frac{R\omega_r}{Q} \\ \frac{\Gamma}{\overline{\omega}_r} &= \frac{1}{\sqrt{4Q^2 - 1}} \\ \overline{\omega}_r &= \omega_r\sqrt{1 - \frac{1}{4Q^2}} \end{aligned} \tag{20}$$

A qualitative behaviour of the wake function is shown in Fig. 6.

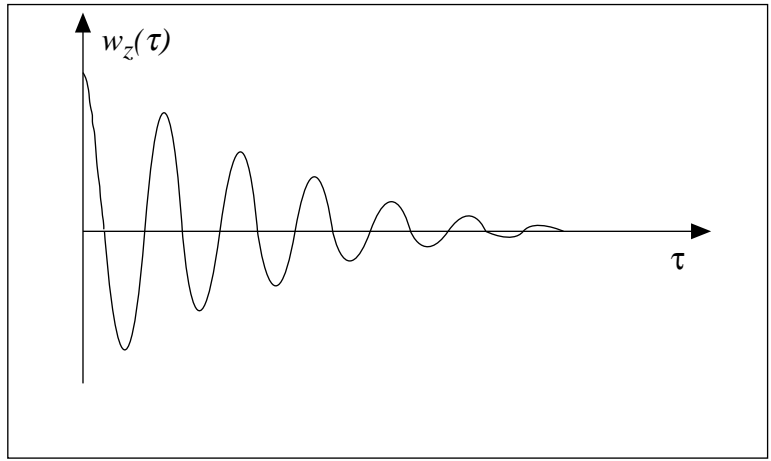

Fig. 6 - Wake function of a resonant mode

The loss factor, according to the definition (4), can be computed as the energy lost by the unit charge after its passage through the cavity. Applying the energy conservation law, we can obtain the energy lost by the charge $q_1$ as the e.m. energy initially stored in the capacitor. We get:

$$k = \frac{1}{2C} = \frac{w(\tau \to 0)}{2} \tag{21}$$

which satisfies the beam loading theorem.

In terms of the merit factor Q we get:

$$k = \frac{R\omega_r}{2Q} \tag{22}$$

### 2.3 Longitudinal wake function and loss factor of a bunch.

The wake function defined in Eq.(5), being generated by a point charge, is a Green function and allows to compute the wake produced by any bunch distribution. Let us consider now a bunch of particles moving on a trajectory parallel to the axis, at a distance $r_1$, with a longitudinal time distribution function $i_b(\tau)$ such that:

$$q_1 = \int_{-\infty}^{+\infty} i_b(\tau) \, d\tau \tag{23}$$

The wake function produced by the bunch distribution at a point with time delay $\tau$ is simply given by the convolution of the Green function over the bunch distribution. We remind that, in practice, the convolution integral is obtained by applying the superposition principle. We split the distribution into an infinite number of infinitesimal slices and sum up their wake contributions at the point $\tau$. According to the definitions given so far, the energy lost by a trailing charge $q$ because of the wake produced by the slice at $\tau'$ is:

$$dU(\mathbf{r}, \tau - \tau') = q \, i_b(\tau') w_z(\mathbf{r}, \tau - \tau') d\tau' \tag{24}$$

summing up all the effects we get the wake function of a bunch distribution as:

$$W_z(\mathbf{r}, \tau) = \frac{U(\mathbf{r}, \tau)}{q_1 q} = \frac{1}{q_1} \int_{-\infty}^{\infty} i_b(\tau') w_z(\mathbf{r}, \tau - \tau') d\tau' \tag{25}$$

For a bunch traveling with velocity "c", because of the causality, the above folding integral has the observation point "$\tau$" as uppermost limit.

Once the bunch wake function $W_z(r, \tau)$ is known, it is straightforward to derive the loss factor of the charge distribution by applying again the superposition principle, we get:

$$K(r) = \frac{U(r)}{q_1^2} = \frac{1}{q_1} \int_{-\infty}^{\infty} W_z(r, \tau) \, i_b(\tau) \, d\tau \tag{26}$$

which depends on the transverse displacement of the bunch.

2.3.1 Example: rectangular bunch distribution exciting a single HOM.

Let us consider a bunch distribution with a simple rectangular shape on the axis at r=0.

$$i_b(\tau) = \frac{q_1}{2T}[H(\tau + T) - H(\tau - T)] \tag{27}$$

and compute the wake function of such a charge distribution assuming that it excites a single HOM in a r.f. cavity. Further, let us assume that the factor Q is so high that, in the range of interest, the impulsive wake function can be approximated by:

$$w_z(\tau) = w_o \cos(\omega_r \tau) H(\tau) \tag{28}$$

By using the folding integral (25) we get two expressions of the bunch wake for $\tau$ inside and outside the distribution. Inside the charge distribution, i.e. for $-T < \tau < T$, we get:

$$W_z(\tau) = \frac{w_o}{2} \frac{\sin[\omega_r(\tau + T)]}{\omega_r T} H(\tau + T) \tag{29}$$

It is worth noting that in the limit $T \to 0$, the rectangular distribution becomes an impulsive function $i_b(\tau) = q_1 \delta(\tau)$ and the bunch wake $W_z(\tau) \to w_z(\tau)$. In particular it is interesting to see that:

$$\lim_{T \to 0} W_z(0) = \frac{w_o}{2} \tag{30}$$

Namely, looking at the center of the bunch, one finds that the wake function approaches with continuity the limit value (7).

The bunch loss factor is obtained from Eq.(26) which gives:

$$K = \frac{w_o}{2}\left(\frac{\sin(\omega_r T)}{\omega_r T}\right)^2 \tag{31}$$

The "point charge" loss factor is derived from the above expression in the limit $T \to 0$:

$$k = \lim_{T \to 0} K = \frac{w_o}{2} \qquad (32)$$

Therefore, when we consider any bunch distribution, the somewhat "artificious" arguments presented in Sec. 2.2 are unnecessary, since the loss factor can be computed strightforwadly from the bunch wake which turns out to be continuous, even in the "point charge" limit.

Finally we find externally to the distribution, i.e. for $\tau \geq T$:

$$W_z(\tau) = w_o \frac{\sin(\omega_r T) \cos(\omega_r \tau)}{\omega_r T} H(\tau - T) \qquad (33)$$

It is interesting to note that outside the distribution, the limit for $T \to 0$ and $\tau \to 0$ of Eq. (33) gives $w_o$.

**2.4 Loss factor and Poynting Vector**

The bunch wake function has been defined as the energy loss by a bunch crossing a given structure. We already said that the non-consistency due to the constant velocity of the bunch can be avoided assuming an external force acting on the bunch (for instance related to an electric external potential). Since the kinetic energy of the bunch is constant (constant velocity) the work done by the external force has to be equal to the energy loss, according to the energy conservation law. However, it is well known that, any electromagnetic energy loss can be computed as the flux of the Poynting vector over a closed surface surrounding the sources of the fields.

The Poynting theorem states that the electromagnetic energy $U_{em}$ stored in a volume V limited by the surface S can change because of homic losses and electromagnetic radiation:

$$\frac{\partial U_{em}}{\partial t} = -\int_S \boldsymbol{P} \cdot \hat{n} \, dS + \int_V (\boldsymbol{E} \cdot \boldsymbol{J}) \, dV \qquad (34)$$

where $\hat{n}$ is the unity normal to the surface S, $\boldsymbol{J}$ is the current density, $\boldsymbol{E}$ the electric field and $\boldsymbol{P}$ is the Poynting vector defined as:

$$\boldsymbol{P} = \frac{1}{\mu} \boldsymbol{E} \times \boldsymbol{B} \qquad (35)$$

Let us consider now a single charge moving on the axis of a given structure. The current density is given by:

$$J(r,z,t) = q_1 v \frac{\delta(r)}{2\pi r} \delta(z-vt) \tag{36}$$

We choose as surface S a cylinder of infinitesimal radius around the charge trajectory. Integrating Eq. (34) with respect to the time from $-\infty$ to $+\infty$, and noting that in the volume V $U_{em}(t=-\infty) = U_{em}(t=\infty)$, we get:

$$\int_{-\infty}^{\infty} dt \int_V (E \cdot J)\, dV = \int_{-\infty}^{\infty} dt \int_S P \cdot \hat{n}\, dS \tag{37}$$

Making use of (2),(4) and (36) we get for the loss factor:

$$k = \frac{-1}{q_1} \int_{-\infty}^{\infty} E_z\left(z, t=\frac{z}{v}\right) dz = \frac{-1}{q_1} \int_{-\infty}^{\infty} dt \int_S P \cdot \hat{n}\, dS \tag{38}$$

**2.5 The synchronous fields.**

When a bunch crosses the various elements installed on the beam pipe, it excites secondary fields because of induction effects and diffraction phenomena. Some of these fields are localized around the bunch, as for example the space charge or the resistive wall fields, others are localized in resonant structures like the r.f. cavity, and others, at high frequency can propagate within the beam pipe. All these fields interact with the circulating beam.

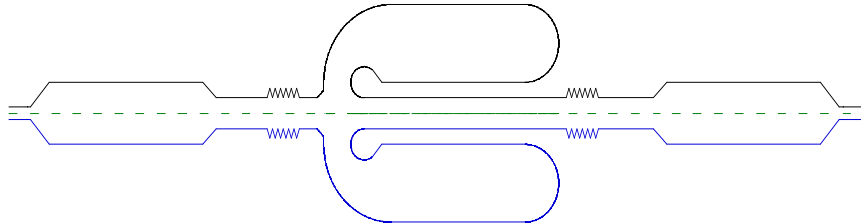

Fig. 7 - DAΦNE Accumulator vacuum chamber (RF cavity, kicker tanks, bellows)

We want to show that this interaction is such that only the fields components synchronous with the charges can change the charges energy. In order to prove this statement it is convenient to express the longitudinal electric field in terms of waves propagating in the z-direction:

$$E_z(z,t) = \frac{1}{2\pi} \int_{-\infty}^{\infty} d\omega \int_{-\infty}^{\infty} d\kappa \, \tilde{E}_z(\omega, \kappa) e^{j(\omega t - \kappa z)} \tag{39}$$

where we have omitted the explicit dependence on $(r, r_1, z_1)$.

The longitudinal electric field is given by a superposition of longitudinal waves having any phase velocity, since $\omega$ and $\kappa$ can vary from $-\infty$ to $\infty$. Among these waves only those having the same phase velocity of the charge can contribute to the energy gain and therefore to the wake function. In fact let us put the field expression (39) into the wake function definition (5), we get:

$$w_z(\tau) = \frac{-1}{q_1} \int_{-\infty}^{\infty} E_z\left(z, t = \frac{z}{v} + \tau\right) dz = \frac{-1}{2\pi q_1} \int_{-\infty}^{\infty} d\omega \, e^{j\omega\tau} \int_{-\infty}^{\infty} d\kappa \, \tilde{E}_z(\omega, \kappa) \int_{-\infty}^{\infty} e^{-jz(\kappa - \kappa_o)} dz \tag{40}$$

with $\kappa_o = \frac{\omega}{v}$. We recognize the impulsive function:

$$\frac{1}{2\pi} \int_{-\infty}^{\infty} e^{-jz(\kappa - \kappa_o)} dz = \delta(\kappa - \kappa_o) \tag{41}$$

that allows to get the following simple expression:

$$w_z(\tau) = \frac{-1}{q_1} \int_{-\infty}^{\infty} \tilde{E}_z(\kappa = \kappa_o, \omega) e^{j\omega\tau} d\omega \tag{42}$$

wherein it is apparent that only those components of the fields propagating with the same phase velocity of the charge can produce a "surfing" effect. All the others in average do not contribute.

The result found above deserve a further investigation. In fact we wonder what would happen if, instead of an infinite structure one would consider a pipe with finite length, say L. It is easy to see that integration between -L/2 and +L/2 does not give an impulsive function, but:

$$\frac{1}{2\pi} \int_{-L/2}^{L/2} e^{-jz(\kappa-\kappa_o)} dz = \frac{L}{2\pi} \frac{\sin\left[(\kappa-\kappa_o)\frac{L}{2}\right]}{(\kappa-\kappa_o)\frac{L}{2}} \tag{43}$$

which becomes again an impulsive function when $L \to \infty$. For a finite length L, the "sinc" function (sin(x)/x) has a maximum at $\kappa_o = \kappa$, and the first zero at $\kappa_o = \kappa \pm 2\pi/L$.

For long wavelength, the fields do not propagate being stored within a given device (e.g. the cavity HOMs). The actual integration path is therefore confined to a limited region, the fields being evanescently zero above the pipe cut-off. On the other hand, for short wavelengths fields propagate into the beam pipe. There is a contribution of those harmonics that do not perfectly average to zero their effect on the beam. However, according to Eq. (43) at high frequencies this contribution is small, so that we can consider an infinite pipe instead of a finite one, simplifying the calculation of the wake.

## 2.6 Expansion of the longitudinal wake in cylindrical symmetry

So far we have considered the case of general boundaries, assuming the two charges moving on any trajectory parallel to the axis. We have already mentioned that in general there is no restriction on the transverse position of both charges. For simplicity we now consider that the trajectories are parallel to the axis of a structure with cylindrical symmetry as shown in Fig. 2. Let $(r_1, \phi_1 = 0, z_1)$ be the coordinates of the leading charge and $(r, \phi, z)$ those of the trailing one. The density charge $q_1$ can be represented as a superposition of multipole moments in cylindrical coordinates:

$$\rho_1 = q_1 \frac{\delta(r-r_1)}{r_1} \delta(\phi) \delta(z-z_1) \tag{44}$$

with $z_1 = \beta c \tau$. Exploiting the azimuthal periodicity we can write:

$$\rho_1 = \frac{q_1}{2\pi} \frac{\delta(r-r_1)}{r_1} \delta(z-z_1) \sum_{m=0}^{\infty} \alpha_m \cos(m\phi) \tag{45}$$

with:

$$\alpha_m = \begin{cases} 1, & m=0 \\ 2, & m \neq 0 \end{cases} \tag{46}$$

According to the above expression the charge can be thought of as a superposition of charged rings with angular dependence $cos(m\phi)$. It easy to see for instance that the monopolar term with m=0 describes a charged ring of radius $r_1$ with uniform density. In cylindrical coordinates the e.m. fields created by the distribution (45) can be derived as sum of multipole terms as well, showing therefore the same angular dependence. For each term we can compute the effect of the longitudinal force. The resulting wake function will show the following form:

$$w_z(\boldsymbol{r},\boldsymbol{r}_1;\tau) = \sum_{m=0}^{\infty} w_{z,m}(\boldsymbol{r},\boldsymbol{r}_1;\tau) \tag{47}$$
$$w_{z,m}(\boldsymbol{r},\boldsymbol{r}_1;\tau) = \overline{w}_{z,m}(r,r_1;\tau)cos(m\phi)$$

## 2.7 Radial expansion of the wake function in the limit $\boxed{\gamma \rightarrow \infty}$.

The e.m. fields produced by the traveling charge in a vacuum chamber are derived from Maxwell equation imposing the boundary conditions at the pipe walls:

$$\nabla \times \boldsymbol{B} = \mu_o \boldsymbol{J} + \frac{1}{c^2}\frac{\partial \boldsymbol{E}}{\partial t} \tag{48}$$

$$\nabla \times \boldsymbol{E} = -\frac{\partial \boldsymbol{B}}{\partial t} \tag{49}$$

$$\nabla \cdot \boldsymbol{E} = -\frac{\rho}{\varepsilon_o} \tag{50}$$

$$\boldsymbol{J} = \rho \boldsymbol{v} \tag{51}$$

The longitudinal electric field can be thought of as produced by the current sources (the bunch) and by the currents induced at the walls. Considering only the induced fields, it can be shown that the fourier component synchronous with the charges is a solution of the following equation [6,7]:

$$\nabla_{\perp}^2 \tilde{E}_z - \left(\frac{\omega}{\beta c \gamma}\right)^2 \tilde{E}_z = 0 \tag{52}$$

In the limit $\gamma \rightarrow \infty$, i.e. in the case of ultra relativistic charges, we have:

$$\nabla_{\perp}^2 \tilde{E}_z = 0 \tag{53}$$

Solved in cylindrical symmetry, the above equation gives the following radial dependence of the wake function [4,6,7]:

$$\overline{w}_{z,m}(r,r_1;\tau) = r^m r_1^m \,\overline{\overline{w}}_{z,m}(\tau) \tag{54}$$

The monopole term m=0, does not depend on the radial position of the charges. This result, applied only to ultra relativistic charges, allows to simplify the evaluation of the wake function by choosing a suitable integration path. Numerical codes [1,25,57,58] computing the longitudinal monopole wake function of charges with β=1, in structure with cylindrical symmetry, perform the integration along trajectories at the radius of the beam pipe. Since the longitudinal electric field vanishes on the pipe surface, the integration is limited to a shorter path.

We want to underline that the expansion (54) concerns only the secondary fields induced by the beam. The primary fields produce the so called space charge wake effects that show a different radial dependence, (Sec. 6.3).

## 2.8 Wake function in accelerator rings

In the case of circular machines, the longitudinal position of the charge is given by the coordinate θ. We compute the wake function by averaging the azimuthal electric field over a revolution period $T_o$:

$$w(\tau) = -2\pi R \langle E_\theta(\theta, t+\tau) \rangle_{T_o} \tag{55}$$

Due to the intrinsic periodicity of the e.m. problem, we can expand the longitudinal electric field of a single charge as:

$$E_\theta(\theta,t) = \int_{-\infty}^{\infty} d\omega\, e^{j\omega t} \sum_{n=-\infty}^{\infty} \tilde{E}_z(n,\omega) e^{-jn\theta} \tag{56}$$

which substituted in (55) gives:

$$w(\tau) = -R \int_{-\pi}^{\pi} d\theta \int_{-\infty}^{\infty} d\omega\, e^{j\omega\tau} \sum_{n=-\infty}^{\infty} \tilde{E}_\theta(n,\omega) e^{-j\theta(n-\omega/\omega_o)} \tag{57}$$

The charge itself can be thought of as a train of charges with a beam current:

$$i_b(\tau) = q_1 \sum_{k=-\infty}^{\infty} \delta(\tau - kT_o)$$

Making use of (25) and (55), we have:

$$W(\tau) = \frac{1}{q_1} \int_{-\infty}^{\infty} i_b(\tau') w(\tau - \tau') d\tau' = \sum_{k=-\infty}^{\infty} w\left(\tau - \frac{2\pi k}{\omega_o}\right) \tag{58}$$

$$W(\tau) = -R \int_{-\pi}^{\pi} d\theta \int_{-\infty}^{\infty} d\omega \sum_{k=-\infty}^{\infty} e^{j\omega(\tau - 2\pi k/\omega_o)} \sum_{n=-\infty}^{\infty} \tilde{E}_\theta(n, \omega) e^{-j\theta(n - \omega/\omega_o)}$$

which, after some mathematics, becomes:

$$W(\tau) = -\frac{2\pi R}{q_1} \sum_{n=-\infty}^{\infty} \tilde{E}_\theta(n, n\omega_o) e^{jn\omega_o \tau} \tag{59}$$

## 3. LONGITUDINAL COUPLING IMPEDANCE

### 3.1 Definitions and properties

In the frequency domain we compute the spectrum of the point charge wake function as:

$$\int_{-\infty}^{\infty} w_z(\mathbf{r}, \mathbf{r}_1; \tau) e^{-j\omega\tau} d\tau \equiv Z(\mathbf{r}, \mathbf{r}_1; \omega) \tag{60}$$

Which being measured in *Ohms* units is called *Coupling Impedance*. Historically, the coupling impedance concept was introduced in the early studies of the instabilities arising in the ISR at CERN [8].

The wake function is derived from the impedance by inverting the Fourier integral:

$$w_z(\mathbf{r}, \mathbf{r}_1; \tau) = \frac{1}{2\pi} \int_{-\infty}^{\infty} Z(\mathbf{r}, \mathbf{r}_1; \omega) e^{j\omega\tau} d\omega \tag{61}$$

In the following we shall omit, for simplicity, the radial dependence. Comparison with Eq.(42) shows that:

$$Z(\omega) = -\frac{2\pi}{q_1} \tilde{E}_z(\kappa = \kappa_o, \omega) \tag{62}$$

The coupling impedance is a complex quantity:

$$Z(\omega) = Z_r(\omega) + j Z_i(\omega) \tag{63}$$

with $Z_r(\omega)$ and $Z_i(\omega)$ even and odd function of $\omega$ respectively. It is easy to prove this property of the impedance reminding that the wake potential $w(\tau)$ is a real function of $\tau$. In fact expanding the exponential in the integral of Eq. (61) we have:

$$w_z(\tau) = \frac{1}{2\pi} \int_{-\infty}^{\infty} [Z_r(\omega)\cos(\omega\tau) - Z_i(\omega)\sin(\omega\tau)] d\omega$$
$$+ \frac{j}{2\pi} \int_{-\infty}^{\infty} [Z_r(\omega)\sin(\omega\tau) + Z_i(\omega)\cos(\omega\tau)] d\omega \tag{64}$$

where the imaginary part vanishes if:

$$\begin{aligned} Z_r(\omega) &= Z_r(-\omega) \\ Z_i(\omega) &= -Z_i(-\omega) \end{aligned} \tag{65}$$

From Eqs. (8,60,64), we recognize that $Z_r(\omega)$ and $-Z_i(\omega)$ are the Fourier transform of $w_z^e(\tau)$ and $w_z^o(\tau)$ respectively:

$$w_z^e(\tau) = \frac{1}{2\pi} \int_{-\infty}^{\infty} Z_r(\omega) \cos(\omega\tau) d\omega$$
$$w_z^o(\tau) = \frac{-1}{2\pi} \int_{-\infty}^{\infty} Z_i(\omega) \sin(\omega\tau) d\omega \tag{66}$$

Furthermore, in the particular case of $\beta = 1$, the wake function has to vanish for $\tau < 0$ where. $w_z^e(\tau) = -w_z^o(\tau)$. In terms of impedances Eq.(10) becomes:

$$\int_{-\infty}^{\infty} Z_r(\omega) \cos(\omega\tau) d\omega = \int_{-\infty}^{\infty} Z_i(\omega) \sin(\omega\tau) d\omega \tag{67}$$

which expresses a general relationship between the real and imaginary part of the impedance. It can be shown that the above relation is equivalent to the Hilbert transform relating the real and imaginary part of a network impedance. In other words the Coupling Impedance defined by Eq.(60) behaves like an usual circuit impedance only when the causality principle applies, namely in the limit case of charges traveling with the velocity of light.

Recalling the relation (7) between loss factor and wake potential for a point charge, we get :

$$k = \frac{w_z(\tau \to 0^+)}{2} = \frac{1}{\pi} \int_0^\infty Z_r(\omega)d\omega \qquad (68)$$

where we recognize that the real part of the impedance is the power spectrum of the energy loss of a unit point charge. In general, the complex impedance can be thought of as the complex power spectrum related to the energy loss.

<u>3.1.1 Example: Impedance of a single HOM in a rf cavity</u>.

Using the wake function expression (17) derived for a single HOM, from the definition (60) we have:

$$Z(\omega) = \frac{\omega_r R}{Q} \int_{-\infty}^{\infty} \left[ \cos(\overline{\omega}_r \tau) - \frac{\Gamma}{\overline{\omega}_r} \sin(\overline{\omega}_r \tau) \right] e^{-(j\omega + \Gamma)\tau} d\tau \qquad (69)$$

$$Z(\omega) = \frac{R}{1 + jQ\left(\frac{\omega}{\omega_r} - \frac{\omega_r}{\omega}\right)} \qquad (70)$$

It is easy to verify that the above impedance satisfies the properties (65) and (67).

**3.2 Bunch losses and wake function from the impedance**

Consider a bunch radially displaced and with a charge distribution $i_b(\tau)$, whose Fourier spectrum is $I(\omega)$. The total bunch wake function $W_z(\boldsymbol{r};\tau)$ and loss factor can be expressed in terms of $Z(\omega)$ by transforming the integrals (25) and (26), obtaining:

$$W_z(\boldsymbol{r};\tau) = \frac{1}{2\pi q_1} \int_{-\infty}^{\infty} Z(\boldsymbol{r};\omega) I(\omega)\, e^{j\omega\tau} d\omega \qquad (71)$$

$$K(\boldsymbol{r}) = \frac{1}{\pi q_1^2} \int_0^\infty Z_r(\boldsymbol{r};\omega) |I(\omega)|^2 d\omega \qquad (72)$$

As an example for a bunch with Gaussian distribution:

$$I(\omega) = q_1 \, e^{-\frac{(\omega \sigma_\tau)^2}{2}} \tag{73}$$

the loss factor is given by:

$$K(r) = \frac{1}{\pi} \int_0^\infty Z_r(r;\omega) \, e^{-(\omega \sigma_\tau)^2} d\omega \tag{74}$$

It is apparent that the loss factor is, in general, a function of the r.m.s. length of the bunch distribution. It is interesting to note that there exists a general relation, useful in the measurements, between the frequency dependence of the impedance and the dependence of the loss factor on the bunch length. For a Gaussian bunch:

$$Z_r(\omega) \propto \omega^a \Leftrightarrow K \propto \sigma_\tau^{-(a+1)} \tag{75}$$

### 3.3 Multipole longitudinal impedance for cylindrical symmetry

In Sec. 2.6 we have seen that the wake function created by a charge on a trajectory parallel to the axis of a device with cylindrical symmetry, can be expanded into a sum of multipolar terms. The wake expansion used in the impedance definition allows to express also the impedance as a multipole expansion:

$$Z(r, r_1, \phi; \omega) = \sum_{m=0}^{\infty} Z_m(r, r_1, \phi; \omega) = \sum_{m=0}^{\infty} \bar{Z}_m(r, r_1; \omega) \cos(m\phi) \tag{76}$$

For ultra relativistic charges the radial dependence of the wake function is known, according to Eq.(54), we have:

$$\bar{Z}_m(r, r_1; \omega) = r^m r_1^m \bar{\bar{Z}}_m(\omega) \tag{77}$$

where $\bar{\bar{Z}}_m(\omega)$ has dimensions $\Omega/\text{m}^{2m}$.

## 4    TRANSVERSE WAKE FUNCTION

### 4.1 Transverse wake function and loss factor of a point charge

Let us consider now the leading charge transversely displaced with respect to the axis as shown in Fig. 8. The charge excites in the structure electromagnetic fields which can be expanded in their multipole components (dipole, quadrupole, sextupole etc.) in the transverse plane. For small transverse displacements the dipole term is of course dominant.

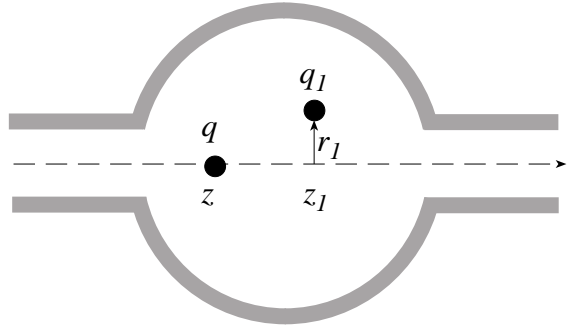

Fig. 8 - The leading charge is transversely displaced

The trailing charge $q$ experiences a Lorentz force which has longitudinal and transverse components. Therefore, it is subject to a transverse momentum kick given by:

$$M_{21}(r,r_1;\tau) = \int_{-\infty}^{\infty} F_\perp(r,z,r_1,z_1\,;\,t)\,dz\,, \quad t = \frac{z_1}{v} + \tau \quad (78)$$

the integration, as for the longitudinal case is assumed over an infinite distance. The above momentum kick, measured in *Newton meter* [Nm], depends on the pipe shape and on the transverse position of both charges. In general, the transverse kick is not parallel to the displacement of the leading charge. In fact an horizontal displacement can lead to both vertical and horizontal kicks, and the same generally happens for a vertical displacement. Only in the case of cylindrical symmetry, the two transverse directions are de-coupled for a beam on the axis. The transverse kick per unit of both charges, measured in *volt/coulomb* [V/C], defines the transverse wake function:

$$w_\perp(r,r_1;\tau) = \frac{M_{21}(r,r_1;\tau)}{q_1 q} \quad [\text{V/C}] \quad (79)$$

Analogously to the longitudinal case we find useful to define the dipole transverse loss factor as the amplitude of the transverse momentum kick given to the charge by its own wake per unit charge:

$$k_\perp(r_1) = \frac{M_{11}(r_1)}{q_1^2} \qquad [\text{V/C}] \tag{80}$$

Usually the dipole component of the transverse kick is the dominant term for ultra relativistic charges. This term is proportional to the displacement of the charge $q_1$. For this particular case we define the *transverse dipole wake function* as the transverse wake per unit of transverse displacement:

$$w'_\perp(r, r_1; \tau) = \frac{w_\perp(r, r_1; \tau)}{r_1} \qquad [\text{V/Cm}] \tag{80}$$

and a transverse loss factor:

$$k'_\perp(r_1) = \frac{M_{11}(r_1)}{q_1^2 r_1} \qquad [\text{V/Cm}] \tag{81}$$

## 4.2 Transverse wake function and loss factor of a bunch

The transverse wake potential produced by a continuous bunch distribution, transversely displaced by $r$, can be obtained by applying the superposition principle; we get:

$$W_\perp(r; \tau) = \frac{1}{q_1} \int_{-\infty}^{\infty} w_\perp(r; \tau - \tau') \, i_b(\tau') d\tau' \qquad [\text{V/C}] \tag{82}$$

The bunch transverse loss factor [V/C] is:

$$K_\perp(r) = \frac{1}{q_1} \int_{-\infty}^{\infty} W_\perp(r; \tau) \, i_b(\tau) d\tau \tag{83}$$

the transverse wake and loss factor per unit displacement are:

$$W'_\perp(r; \tau) = \frac{W_\perp(r; \tau)}{r} \tag{84}$$

$$K'_\perp(r) = \frac{K_\perp(r)}{r} \tag{85}$$

measured in *volt/(coulomb meter)*.

## 4.3 Relationship between longitudinal and transverse wake functions

Let us consider, for simplicity, a charge which moving with constant velocity **v** along the z-axis, through an e.m. field. It will experience a e.m. force with components:

$$F_z = q E_z$$
$$F_r = q(E_r - vB_\phi) \qquad (86)$$
$$F_\phi = q(E_\phi + vB_r)$$

The Maxwell equations give:

$$\frac{\partial E_z}{\partial r} = \frac{\partial E_r}{\partial z} + \frac{\partial B_\phi}{\partial t}$$
$$\frac{1}{r}\frac{\partial E_z}{\partial \phi} = \frac{\partial E_\phi}{\partial z} - \frac{\partial B_r}{\partial t} \qquad (87)$$

from the above relationships, in a moving frame $\zeta = z - vt$, we derive:

$$\nabla_\perp F_z = \frac{\partial \mathbf{F}_\perp}{\partial \zeta} \qquad (88)$$

The moving frame has as origin of the axis the position of the leading charge ($z=vt \rightarrow \zeta=0$), while on the trailing charge we have $\zeta=-v\tau$. Since we are considering the force on the trailing charge, derivation with respect to $\zeta$ can be substituted with derivation with respect to $\tau$:

$$-\frac{1}{v}\frac{\partial}{\partial \tau}\mathbf{w}_\perp(\mathbf{r},\mathbf{r}_1;\tau) = \nabla_{\perp,r} w_z(\mathbf{r},\mathbf{r}_1;\tau) \qquad [\text{V/Cm}] \qquad (89)$$

The transverse operator $\nabla_\perp$ applies on $\mathbf{r}$, the transverse coordinates of the trailing charge. The above relation is often referred to as "Panofsky-Wenzel" theorem [9].

If the leading charge is slightly displaced from the axis, we can expand the rhs of (89) retaining only the first order term [10]:

$$w_z(\mathbf{r},\mathbf{r}_1;\tau) \approx w_z(\mathbf{r},0;\tau) + \left[\nabla_{\perp,r_1} w_z(\mathbf{r},\mathbf{r}_1;\tau)\right]_{\mathbf{r}_1=0} \cdot \mathbf{r}_1 + O(r_1)^2 \qquad (90)$$

where $\nabla_{\perp,r_1}$ is the gradient operator acting on the transverse coordinates $\mathbf{r}_1$ of the leading charge. From (89) we get:

$$-\frac{1}{v}\frac{\partial}{\partial \tau}\mathbf{w}_\perp(\mathbf{r},\mathbf{r}_1;\tau) = \nabla_{\perp,r}\left\{w_z(\mathbf{r},0;\tau) + \left[\nabla_{\perp,r_1} w_z(\mathbf{r},\mathbf{r}_1;\tau)\right]_{\mathbf{r}_1=0} \cdot \mathbf{r}_1\right\} \qquad (91)$$

The first term in the brackets is a "monopole" contribution to the transverse impedance which disappears for a particular symmetric geometry (circular, rectangular, elliptic). The latter is the dipole transverse impedance, which, in the linear approximation, is obtained from the longitudinal wake expression $w_z(r,r_1;\tau)$ by applying twice the transverse gradient operator to $r_1$ and $r$:

$$-\frac{1}{v}\frac{\partial}{\partial \tau}w_\perp(r,r_1;\tau) = \nabla_{\perp,r}\left[\nabla_{\perp,r_1}w_z(r,r_1;\tau)\right]_{r_1=0} \cdot r_1 \qquad (92)$$

For instance, in Cartesian and cylindrical coordinates, the transverse operator becomes:

$$\nabla_{\perp,r}\nabla_{\perp,r_1} = \begin{vmatrix} \dfrac{\partial^2}{\partial x \partial x_1} & \dfrac{\partial^2}{\partial x \partial y_1} \\ \dfrac{\partial^2}{\partial y \partial x_1} & \dfrac{\partial^2}{\partial y \partial y_1} \end{vmatrix} \quad \nabla_{\perp,r}\nabla_{\perp,r_1} = \begin{vmatrix} \dfrac{\partial^2}{\partial r \partial r_1} & \dfrac{1}{r_1}\dfrac{\partial^2}{\partial r \partial \phi_1} \\ \dfrac{1}{r}\dfrac{\partial^2}{\partial \phi \partial r_1} & \dfrac{1}{rr_1}\dfrac{\partial^2}{\partial \phi \partial \phi_1} \end{vmatrix} \qquad (93)$$

It is worth noting that in general, after the application of the matrix (93) to the vector $r_1$, the transverse dipole wake is not necessarily directed along the offset of the driving charge.

**4.4 Mode expansion in cylindrical symmetry**

As for the longitudinal case, the transverse wake can be expresses as superposition of multipoles terms [11]:

$$w_\perp(r,r_1;\tau) = \sum_{m=0}^{\infty} w_{\perp,m}(r,r_1;\tau) \qquad (94)$$

For $\gamma \to \infty$, making use of the expressions (47),(54) and (89), we get:

$$\frac{\partial}{\partial \tau}w_{\perp,m}(r,r_1;\tau) = -c\,m\,\overline{\overline{w}}_{z,m}(\tau)\,r^{m-1}r_1^m\left\{\cos(m\phi)\hat{r} - \sin(m\phi)\hat{\phi}\right\} \qquad (95)$$

The transverse dipole term m=1 is proportional to the transverse displacement of the leading charge while it does not depend on the transverse position of the trailing one. It is easy to show that in cylindrical coordinates the dipole transverse force is directed along the offset of the leading charge:

$$\frac{\partial}{\partial \tau}w_{\perp,1}(r,r_1;\tau) = -c\,\overline{\overline{w}}_{z,1}(\tau)\,r_1 \qquad (96)$$

where, we remind, $\overline{\overline{w}}_{z,1}$ is the amplitude of the dipole longitudinal wake measured in V/(C m$^2$). The same result is obtained by applying Eq. (92).

## 5. TRANSVERSE COUPLING IMPEDANCE

### 5.1 Definitions and properties

The Fourier transform of the transverse wake function in the frequency domain times the imaginary unity defines the transverse coupling impedance:

$$j \int_{-\infty}^{\infty} w_{\perp}(r, r_2; \tau) \, e^{-j\omega\tau} d\tau \equiv Z_{\perp}(r, r_2; \omega) \quad [\Omega] \tag{97}$$

Historically, the imaginary constant was introduced in order to make the transverse impedance to play the same role as the longitudinal one in the beam stability theory. Since the transverse dynamics is dominated by the dipole transverse wake, we can define the transverse dipole impedance normalized to $r_1$ as:

$$Z'_{\perp}(r_1, r_2; \omega) = \frac{Z_{\perp}(r_1, r_2; \omega)}{r_1} \quad [\Omega/m] \tag{98}$$

it has *ohms/meter* units. Conversely, the transverse wake is obtained from the inverse Fourier transform of the transverse impedance:

$$w_{\perp}(r, r_2; \tau) = \frac{j}{2\pi} \int_{-\infty}^{\infty} Z_{\perp}(r, r_2; \omega) \, e^{j\omega\tau} d\omega \tag{99}$$

### 5.2 Relationship between longitudinal and transverse impedances

The Fourier transform of (89) gives the dipole transverse impedance on terms of the longitudinal one:

$$Z_{\perp}(r, r_1; \omega) = \frac{c}{\omega} \nabla_{\perp} Z(r, r_1; \omega) \quad [\Omega] \tag{100}$$

The transverse dipole impedance for an arbitrary shape, according to Eq. (92) is:

$$Z_{\perp}(r, r_1; \omega) = \frac{c}{\omega} \nabla_{\perp, r} \left[ \nabla_{\perp, r_1} Z(r, r_1; \omega) \right]_{r_1 = 0} \cdot r_1 \tag{101}$$

In cylindrical symmetry, applying (96) or (92) to Eqs. (76) we get:

$$Z_{\perp, 1}(r, r_1; \omega) = \frac{c}{\omega} \bar{\bar{Z}}_1(\omega) \, r_1 \quad [\Omega] \tag{102}$$

# 6. UNIFORM BOUNDARIES

## 6.1 General properties

In this chapter we start the analysis of the wake fields and impedances for some relevant cases: charge in the free space and in a beam pipe with uniform cross section. Fields and potentials for these cases have a common feature: they travel together with the charge. In other words, the fields map does not change during the charge flight, as long as the trajectory is parallel to the pipe axis.

Considering a charge with velocity $\mathbf{v}=\beta c \hat{z}$ we may write:

$$\mathbf{E} = -\text{grad } V + \beta^2 \frac{\partial V}{\partial z}\mathbf{z} = -\frac{1}{\gamma^2}\frac{\partial V}{\partial z}\mathbf{z} - \text{grad}_\perp V \tag{103}$$

where the scalar potential $V(r,\phi,z-vt)$ is the solution of the equation:

$$\nabla_\perp^2 V + \frac{1}{\gamma^2}\frac{\partial^2 V}{\partial z^2} = -\frac{\rho}{\varepsilon} \tag{104}$$

satisfying the boundary conditions. The Laplacian operator $\nabla_\perp^2$ is applied to the transverse coordinates, $\rho$ is the charge density. The longitudinal wake potential per unit length is given by:

$$\frac{\partial w(r,\phi,\tau)}{\partial z} = -\frac{1}{q}E_z(r,\phi,z-vt)\Big|_{t=\frac{z}{v}+\tau} \tag{105}$$

One can see from Eq. (104) that in the ultra relativistic limit $\gamma \to \infty$, fields can be derived in the static approximation.

## 6.2 Relativistic charge in the free space

A point charge moving with constant velocity $v\hat{z}$ in the free space generates fields which are solution of the Maxwell equations (48, 51). The fields can be derived applying the Lorentz transform to the static field created by the charge in the rest frame. Because of the symmetry, we have the fields [12]:

$$\mathbf{E}(r,z,t) = \frac{q}{4\pi\varepsilon_o} \frac{\gamma[\mathbf{r}+(z-vt)\hat{z}]}{[\gamma^2(z-vt)^2+r^2]^{\frac{3}{2}}} \tag{106}$$

$$\mathbf{B}(r,z,t) = \frac{\beta}{c}\hat{n} \times \mathbf{E}(r,z,t) \tag{107}$$

where the $\hat{n}$ vector is directed from the charge to the observation point. The magnetic field has only the azymuthal component $B_\phi$. It is well known that at high energies, because of the relativistic contraction, the fields are mainly confined inside a region with an opening angle $1/\gamma$ and perpendicular to the trajectory. The longitudinal field $E_z$ vanishes as $1/\gamma^2$, while $E_r$ and $B_\phi$ are proportional to $\gamma$.

$$E_z\left(r=0, t=\frac{z}{v}+\tau\right) = \frac{-q}{4\pi\varepsilon_o(\beta\gamma c\tau)^2} \tag{108}$$

$$E_r\left(r, t=\frac{z}{v}\right) = \frac{q}{4\pi\varepsilon_o}\frac{\gamma}{r^2} \tag{109}$$

$$B_\phi\left(r, t=\frac{z}{v}\right) = \frac{qZ_o}{4\pi}\frac{\gamma}{r^2} \tag{110}$$

Because of the fields confinement within an angular region of the order of $1/\gamma$, at a given distance $r$ from the charge the fields can be thought of as generated by a relativistic charge distribution with line density $\lambda$. In the stationary approximation, applying the Gauss law at a cylindrical surface of radius r, we find an effective charge density $\lambda = q\gamma/r$. The singularities at $\tau = 0$ and $r = 0$ that can be removed by considering a charge with longitudinal and radial distribution.

According to the definition (6), since a test charge on the axis would experience a repulsive force independently of its position, the longitudinal wake function per unit length is an odd function of $\tau$. The corresponding impedance is purely imaginary. Because of the lack of interest, we do not derive the explicit expressions of the wake and impedance. However, it is interesting to compute the amount of e.m. energy stored in a region outside a tube of radius $b$:

$$U(r \geq b) = \frac{3\pi}{16}\left(\frac{r_o}{b}\right)\gamma m_o c^2 \tag{111}$$

where $r_o$ is the classic radius and $m_o$ is the rest mass of an electron. The e.m. energy is proportional to the kinetic energy of the charge.

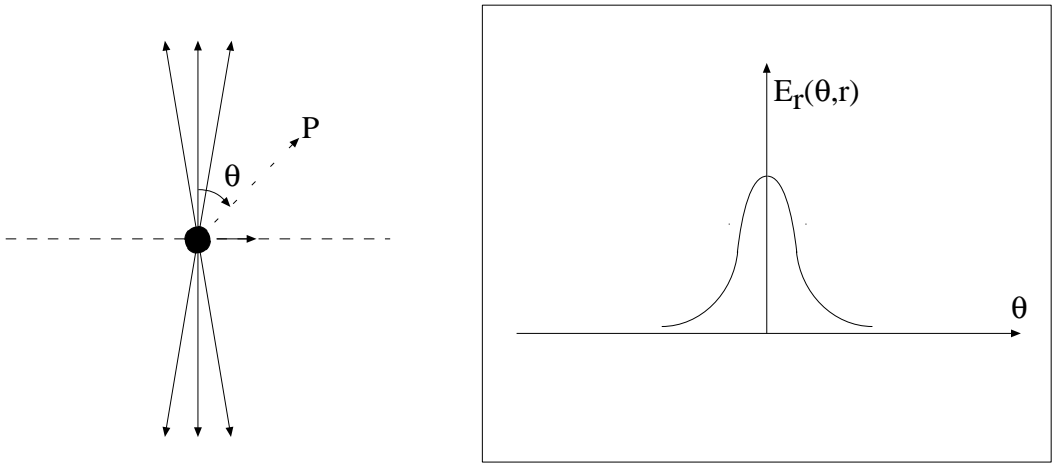

Fig. 9 - Electric field lines of a ultra relativistic charge in the free space
(Qualitative behaviour)

## 6.3 Cylindrical pipe with perfectly conducting walls

The fields produced by a point charge traveling inside a perfectly conducting cylindrical pipe are found from the scalar potential $V(r,\phi,z-vt)$ solution of the Maxwell equation (48,51), with homogeneous boundary conditions at the pipe wall $r=b$ (Fig. 10).

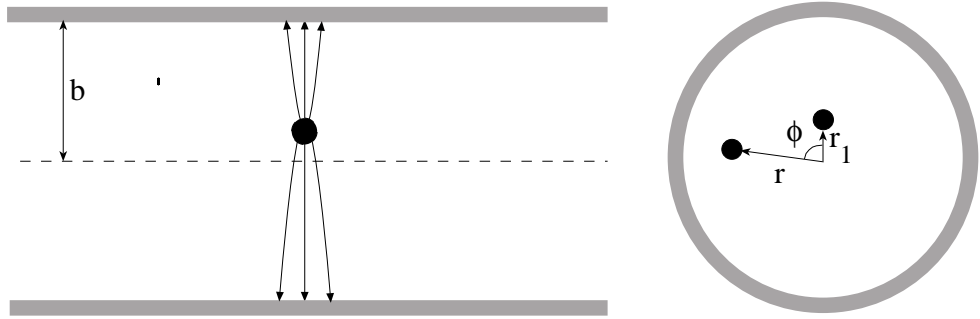

Fig. 10 - Point charge inside a p.c. cylindrical pipe

Using the density charge (44), in cylindrical coordinates we can express the scalar potential as sum of multipole terms [10]:

$$V(r,\phi,z-vt) = \frac{1}{2\pi}\sum_{m=0}^{\infty}\cos(m\phi)\int_{-\infty}^{\infty}\tilde{V}_m(r,r_1,\kappa)e^{-j\kappa(z-vt)}d\kappa \qquad (112)$$

where we made use of the Fourier transform from the z-space to the wave number domain $\kappa$. Each Fourier component $\tilde{V}_m(r,r_1,\kappa)$ is obtained by solving the differential equation (104) and imposing the boundary conditions at the pipe walls [10]. We get:

$$\tilde{V}_m(r,r_1,\kappa) = \frac{q\alpha_m}{2\pi\varepsilon_o}\begin{cases} K_m(\xi r)I_m(\xi r_1) - \dfrac{I_m(\xi r_1)}{I_m(\xi b)}K_m(\xi b)I_m(\xi r); & r \geq r_1 \\ K_m(\xi r_1)I_m(\xi r) - \dfrac{I_m(\xi r_1)}{I_m(\xi b)}K_m(\xi b)I_m(\xi r); & r \leq r_1 \end{cases} \qquad (113)$$

where $\xi = \kappa/\beta\gamma$, and $I_m$, $K_m$ are the modified Bessel functions.

The longitudinal coupling impedance per unit length, using (103) and (105), is given by:

$$\frac{\partial \overline{Z}_m(r,r_1,\omega)}{\partial z} = \frac{-j\omega}{q(c\beta\gamma)^2}\tilde{V}_m\left(r,r_1,\kappa=\frac{\omega}{\beta c}\right) \qquad (114)$$

- 29 -### 6.3.1 Monopole Longitudinal Impedance m=0, r<r1

$$\frac{\partial \bar{Z}_{m=0}}{\partial z} = \frac{-j\omega Z_o}{2\pi c(\beta\gamma)^2}\left[K_o(\xi r_1) - \frac{I_o(\xi r_1)}{I_o(\xi b)}K_o(\xi b)\right]I_o(\xi r) \qquad (115)$$

Where $Z_o = \sqrt{\mu_o/\varepsilon_o} = 1/c\varepsilon_o$ is the impedance of the vacuum. The behaviour of the above term, purely imaginary, as function of $\xi b$ is shown in Fig. 11. We note that in the region $\xi b \ll 1$, the impedance per unit length grows linearly, and does not depend on the radial position of the trailing charge:

$$\frac{\partial \bar{Z}_{m=0}}{\partial z} = \frac{-j\omega Z_o}{2\pi c(\beta\gamma)^2}ln\left(\frac{r_1}{b}\right) \qquad (116)$$

For $\xi b \gg 1$, the impedance shows an exponential roll off.

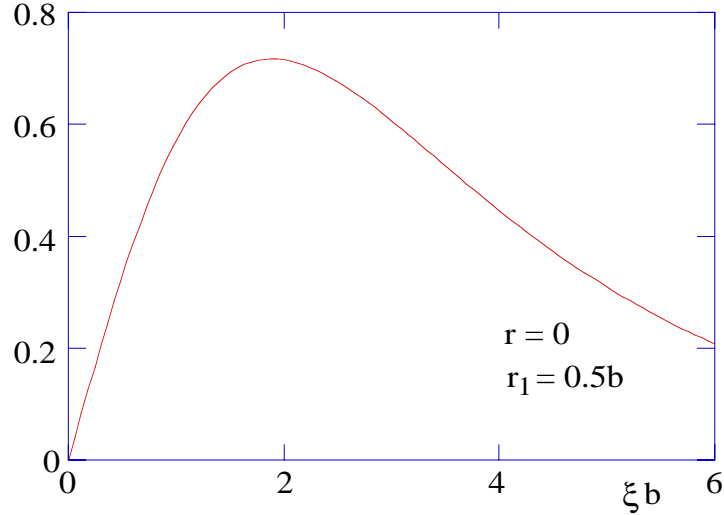

Fig. 11 Monopole space charge impedance versus $\xi b$

It is apparent that the impedance (116) does not satisfy the radial dependence (77) found in the high energy limit. In fact this term is called Space Charge Impedance, and kept distinguished from the forces related to the secondary fields induced at the pipe walls. In the literature, it is usually presented the space charge monopole term due to a disk of radius $a$, centered on the pipe axis. Integrating the impedance expression (116) over the charge distribution $0 \leq r_1 \leq a$, for $\xi b \ll 1$, we get:

$$\frac{\partial \bar{Z}_{m=0}}{\partial z} = \frac{-j\omega Z_o}{4\pi c(\beta\gamma)^2}\left[1 + 2ln\left(\frac{b}{a}\right)\right] \qquad (117)$$

Using the above expression in the inverse Fourier transform, we get in the limit $\gamma \to \infty$ the wake function per unit length:

$$\frac{\partial \overline{w}_{m=0}}{\partial z} = \frac{1}{4\pi\varepsilon_o \gamma^2}\left[1 + 2\ln\left(\frac{b}{a}\right)\right]\frac{\partial}{\partial z}\delta(z-vt) \qquad (118)$$

### 6.3.2 Simple physical approach for $\gamma \to \infty$

We have seen in Sec. 6.1 that at high energies one can solve Maxwell equations in the static approximation. Accordingly we derive the fields produced by a charged cylinder of radius $a$, with longitudinal distribution $\lambda(z-vt)$, moving with light velocity inside a p.c. cylinder (Fig. 12).

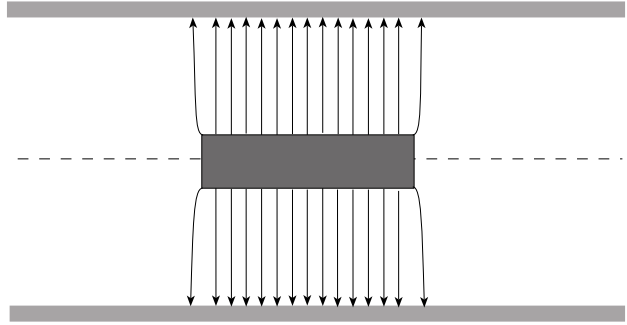

Fig. 12 - On axis cylindrical bunch of radius "a"

For ultra relativistic charges, according to (103) and (104), the fields are derived by applying Gauss and Ampere laws; we get :

$$E_r(r,z-vt) = \begin{cases} \dfrac{\lambda(z-vt)}{2\pi\varepsilon_o}\dfrac{r}{a^2}; & r \leq a \\ \dfrac{\lambda(z-vt)}{2\pi\varepsilon_o}\dfrac{1}{r}; & r \geq a \end{cases} \qquad (119)$$

$$B_\phi(r,z-vt) = \begin{cases} \dfrac{\lambda(z-vt)Z_o}{2\pi}\dfrac{r}{a^2}; & r \leq a \\ \dfrac{\lambda(z-vt)Z_o}{2\pi}\dfrac{1}{r}; & r \geq a \end{cases} \qquad (120)$$

The scalar potential is obtained integrating the radial field (119) from the disk center (r=0) to the pipe radius (r=b); we get:

$$V(r,z-vt) = \frac{\lambda(z-vt)}{4\pi\varepsilon_o}\left\{1 - \left(\frac{r}{a}\right)^2 + 2\ln\left(\frac{b}{a}\right)\right\} \qquad (121)$$

And the wake function per unit length at $r = 0$, $t = \frac{z}{v} + \tau$ becomes:

$$\frac{\partial w_{m=0}}{\partial z} = \frac{1}{q\gamma^2}\frac{\partial V}{\partial z} = \frac{1}{4\pi\varepsilon_o\gamma^2 q}\left[1 + 2\ln\left(\frac{b}{a}\right)\right]\frac{\partial \lambda(z-vt)}{\partial z} \tag{122}$$

which reproduces Eq.(118) for a point charge with density $\lambda(z-vt) = q\delta(z-vt)$.

### 6.3.3 Dipole longitudinal impedance m=1, r<r₁

The dipole impedance per unit length is:

$$\frac{\partial Z_{m=1}}{\partial z} = \frac{-j\omega Z_o}{2\pi c(\beta\gamma)^2}\left[K_1(\xi r_1) - \frac{I_1(\xi r_1)}{I_1(\xi b)}K_1(\xi b)\right]I_1(\xi r)\cos(\phi) \tag{123}$$

In the limit $\xi b \ll 1$ the dipole impedance is proportional to the transverse displacement of the charge:

$$\frac{\partial Z_{m=1}}{\partial z} = \frac{-j\omega Z_o}{4\pi c(\beta\gamma)^2}\left[\frac{1}{r_1^2} - \frac{1}{b^2}\right]rr_1\cos(\phi) \tag{124}$$

### 6.3.4 Dipole transverse impedance $\xi b \ll 1$

Applying the relationship (102) between dipole transverse and longitudinal impedances, and noting that:

$$\nabla_\perp[r\cos(\phi)] = \hat{r}\cos(\phi) - \hat{\phi}\sin(\phi) \equiv \hat{r}_1 \tag{125}$$

we get the transverse dipole impedance per unit length and per unit transverse displacement:

$$\frac{\partial \mathbf{Z}'_{\perp,1}(\omega)}{\partial z} \equiv \frac{1}{r_1}\frac{d\mathbf{Z}_{\perp,1}(\omega)}{dz} = \frac{-jZ_o}{2\pi(\beta\gamma)^2}\left[\frac{1}{r_1^2} - \frac{1}{b^2}\right]\hat{r}_1 \quad [\Omega/\text{m}^2] \tag{126}$$

The same result could be obtained applying Eq.(102), recognizing in (124) the dipole term $\bar{Z}_{m=1}$ introduced in Eq.(76):

$$\frac{\partial \bar{\bar{Z}}_{m=1}}{\partial z} = \frac{-j\omega Z_o}{4\pi c(\beta\gamma)^2}\left[\frac{1}{r_1^2} - \frac{1}{b^2}\right] \tag{127}$$

Notice that according to the standard symbols, also the dipole term can be obtained in terms of radius of a cylindrical beam by putting $r_1 = a$ in Eqs.(124,126,127).

## 6.4 Elliptic pipe with perfectly conducting walls

The impedance expression Eqs.(116),(117), (126) and (127) have been extended to the case of an elliptic pipe [13] in the ultra relativistic limit. An equivalent radius $b_{eq}$ is introduced for both longitudinal and transverse cases as function of the elliptic parameter:

$$\bar{q} = \frac{h-b}{h+b} \quad (128)$$

where $h$ and $b$ are the pipe half-width and half-height respectively. The longitudinal equivalent radius normalized to $b$ is reported in Fig. 13. We see that when $h \gg b$ the curve approaches the parallel plates case with $b_{eq} \cong 4b/\pi$. In Fig.14 the transverse equivalent radius is reported as function of $\bar{q}$ for both horizontal and vertical oscillations.

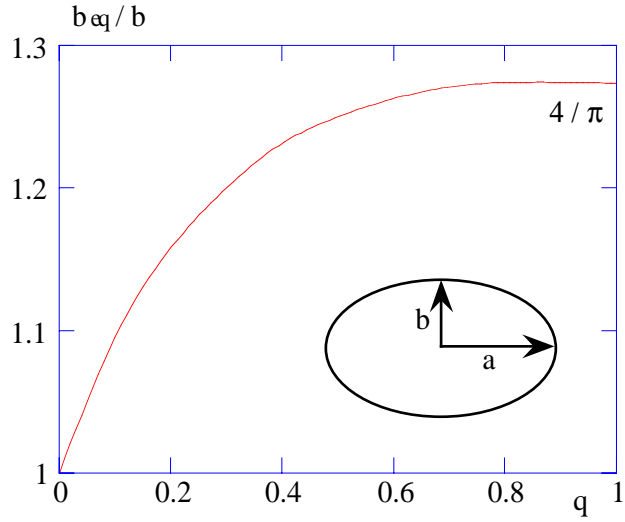

Fig. 13 - Normalized equivalent radius for an elliptic pipe

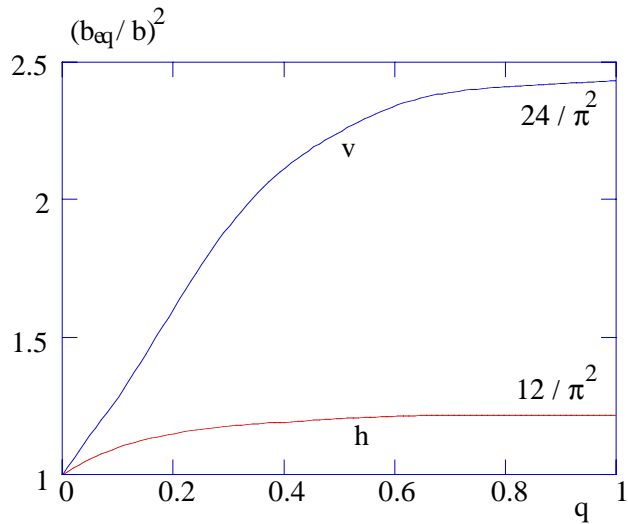

Fig. 13 - Normalized $(b_{eq}/b)^2$ for horizontal (h) and vertical (v) oscillations

**6.5 Pipe with lossy walls**

When consider a pipe with resistive wall of infinite thickness, the Maxwell equations have to be solved both in the pipe space and in the material with finite conductivity $\sigma$ where the fourth Maxwell equation becomes:

$$\nabla \times \boldsymbol{B} = \varepsilon\mu \frac{\partial \boldsymbol{E}}{\partial t} + \mu\sigma \boldsymbol{E} + \mu\rho \boldsymbol{v} \tag{129}$$

Continuity of tangent magnetic field and normal electric field at the wall surface allows to derive the e.m. fields components. The problem of a cylindrical pipe has been solved in the ultrarelativistic limit [14] and for any value of the parameter $\xi b$ [13]. Extension to the elliptic pipe is found in Ref. [15] in the ultra relativistic limit. More recently the impedance for an arbitrary cross section has been developed [16,17] in the same approximation. Here we present the results for the most relevant cases: circular, rectangular and elliptic pipe.

The longitudinal impedance has the general expression:

$$\frac{\partial \bar{Z}_{m=0}}{\partial z} = \frac{1+j}{2\pi b} \sqrt{\frac{\omega Z_o}{2c\sigma}}\, F \tag{130}$$

where F is a form factor depending on the pipe cross section, and $b$ is the half-height of the pipe cross section ($b$ is the radius in the circular case). The inverse Fourier transform of Eq.(130) gives the wake function:

$$\frac{\partial w_{z,m=0}}{\partial z} = -F \frac{1}{4\pi b} \sqrt{\frac{Z_o}{\pi c\sigma}}\, \tau^{-3/2} \tag{131}$$

The above expression, being derived from a static approximation, fails at distances very close to the charge. At very short distances the wake change sign as shown in Fig. 15.

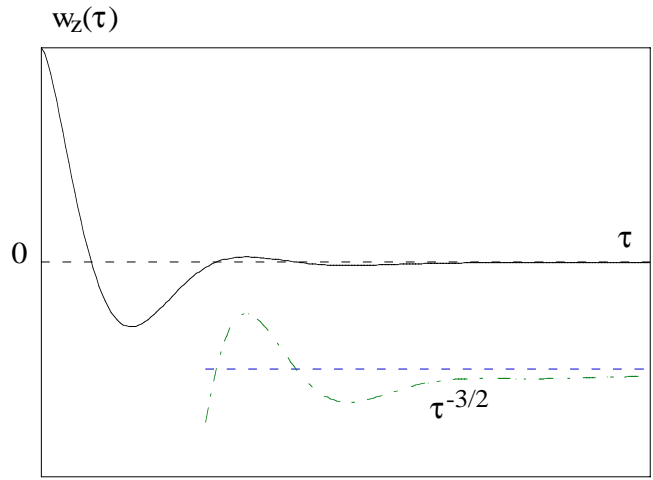

Fig. 15 - Qualitative behaviour of the longitudinal wake of a lossy pipe

The transverse dipole resistive wall impedance is:

$$\frac{\partial \mathbf{Z}'_{\perp,1}}{\partial z} \equiv \frac{1}{r_1}\frac{\partial \mathbf{Z}_{\perp,1}}{\partial z} = F_\perp \frac{1+j}{2\pi b^3} Z_o \delta \hat{r}_1 \qquad [\Omega/\text{m}^2] \qquad (132)$$

where $F_\perp$ is the transverse form factor for vertical and horizontal oscillations. The transverse wake is:

$$\frac{dw'_{\perp,1}(\tau)}{dz} = \frac{1}{r_1}\frac{dw_{\perp,1}(\tau)}{dz} = F_\perp \frac{1}{\pi b^3}\sqrt{\frac{cZ_o}{\pi\sigma}} \, \tau^{-1/2} \qquad [\text{V/m}^2\text{C}] \qquad (133)$$

6.5.1 Circular beam pipe.

For a circular pipe, $F = 1$ and $F_\perp = 1$, the longitudinal monopole impedance is:

$$\frac{\partial \overline{Z}_{m=0}}{\partial z} = \frac{1+j}{2\pi b}\sqrt{\frac{\omega Z_o}{2c\sigma}} \qquad (134)$$

and for the transverse dipole impedance is:

$$\frac{\partial Z_{\perp,1}}{\partial z} = \frac{1+j}{2\pi b^3} Z_o \delta \qquad (135)$$

Simple physical approach:

In the simple case of cylindrical symmetry, according to Sec. 2.4 and 3.1, the longitudinal impedance can be computed as the complex power spectrum related to the energy flowing into the lossy walls. For materials with a high conductivity, the fields inside the pipe are almost the same as in the p.c. case (perturbative approach). In the frequency domain we have:

$$\begin{aligned}\tilde{E}_r(\omega) &= \frac{qZ_o}{2\pi r}\, e^{-jkz} \\ \tilde{H}_\phi(\omega) &= \frac{q}{2\pi r}\, e^{-jkz}\end{aligned} \qquad (136)$$

The continuity conditions at the boundary r=b requires that the magnetic field $\tilde{H}_\phi$ component inside the material surface is the same as outside. Inside the wall the field is sustained by a surface current flowing into the z-direction. The electric field $\tilde{E}_z$ is related to $\tilde{H}_\phi$ by the Leontovich condition:

$$\tilde{E}_z(\omega) = Z_c \tilde{H}_\phi(\omega) \qquad (137)$$

where:

$$Z_c = \sqrt{\frac{j\omega\mu_o}{\sigma}} \tag{138}$$

is the intrinsic impedance of the lossy material. The flux of the Poynting vector at the pipe wall gives:

$$\frac{\partial \overline{Z}_{m=0}}{\partial z} = 2\pi b Z_c \left|\tilde{H}_\phi\right|^2 = \frac{1+j}{2\pi b}\sqrt{\frac{\omega Z_o}{2c\sigma}} \tag{139}$$

### 6.5.2 Rectangular cross section.

For a rectangular beam pipe with half width $h$ and half-height $b$, putting $\lambda = b/h$, the form factor F for the longitudinal impedance is :

$$F(\lambda) = \pi \left[ \sum_{\substack{n=1,\\odd}}^{\infty} \frac{1}{\cosh^2\left(\frac{n\pi}{2\lambda}\right)} + \lambda \sum_{\substack{n=1,\\odd}}^{\infty} \frac{1}{\cosh^2\left(\frac{n\pi\lambda}{2}\right)} \right] \tag{140}$$

The form factor for the dipole transverse impedance in the x-direction:

$$F_x(\lambda) = \frac{\pi^3}{8}\left[ \sum_{\substack{n=1,\\odd}}^{\infty} \frac{n^2}{\sinh^2\left(\frac{n\pi}{2\lambda}\right)} + \lambda^3 \sum_{\substack{n=2,\\even}}^{\infty} \frac{n^2}{\cosh^2\left(\frac{n\pi\lambda}{2}\right)} \right] \tag{141}$$

The function $F_y(\lambda)$ is simply obtained from eq.(141) by moving the factor $\lambda^3$ to the first sum in the brackets:

$$F_y(\lambda) = \frac{\pi^3}{8}\left[ \lambda^3 \sum_{\substack{n=1,\\odd}}^{\infty} \frac{n^2}{\sinh^2\left(\frac{n\pi}{2\lambda}\right)} + \sum_{\substack{n=2,\\even}}^{\infty} \frac{n^2}{\cosh^2\left(\frac{n\pi\lambda}{2}\right)} \right] \tag{142}$$

The behaviour of $F(\lambda)$, $F_x(\lambda)$ and $F_y(\lambda)$ for the rectangular pipe is presented in Fig. 16 as a function of the parameter $\bar{q}$.

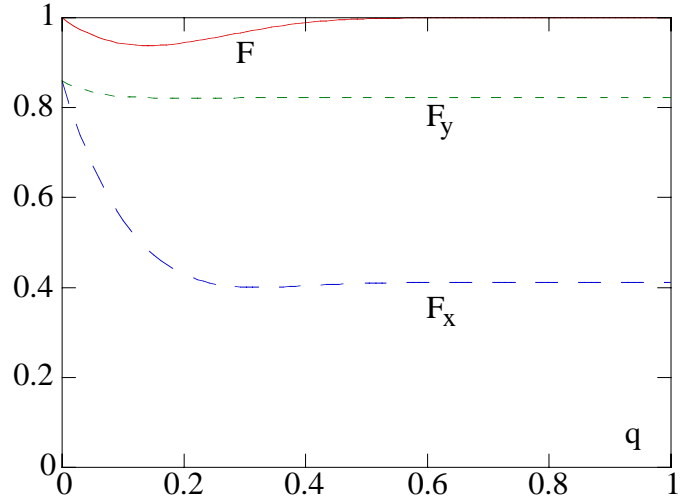

Fig. 16 - Form factors for a rectangular pipe

In the limit case of a pair of parallel plates $\lambda \to 0$, we have:

$$F_o(0) = 1, \quad F_x(0) = \frac{\pi^2}{24}, \quad F_y(0) = \frac{\pi^2}{12}$$

### 6.5.3 Elliptical beam pipe.

For a beam pipe with an elliptical cross-section, major axis $2a$ and minor axis $2b$, the form factor is given as a function of the elliptic parameter $u_o$ related to the parameter $\bar{q}$ by:

$$\bar{q} = e^{-2u_o}$$

We get:

$$F(u_o) = \frac{sinh(u_o)}{2\pi} \int_0^\infty \frac{G(u_o, \alpha) d\alpha}{\sqrt{sinh^2(u_o) + sin^2(\alpha)}} \tag{143}$$

where

$$G(u_o, \alpha) = 2 \sum_{m=-1}^\infty (-1)^m \frac{cos(2m\alpha)}{cosh(2mu_o)} \tag{144}$$

The transverse form factor in the x and y directions is:

$$F_{x,y}(u_o) = \frac{sinh^3(u_o)}{4\pi} \int_0^\infty \frac{G_{x,y}^2(u_o, \alpha) d\alpha}{\sqrt{sinh^2(u_o) + sin^2(\alpha)}} \tag{145}$$

with

$$G_x(u_o, \alpha) = 2\sum_{m=0}^{\infty}(-1)^m(2m+1)\frac{\cos[(2m+1)\alpha]}{\cosh[(2m+1)u_o]} \quad (146)$$

$$G_y(u_o, \alpha) = 2\sum_{m=0}^{\infty}(-1)^m(2m+1)\frac{\sin[(2m+1)\alpha]}{\sinh[(2m+1)u_o]} \quad (147)$$

A graph of the numerical values of $F(u_o)$, $F_{x,y}(u_o)$ for the elliptical pipe is presented in Fig. 17 as a function of the elliptic parameter $\bar{q}$.

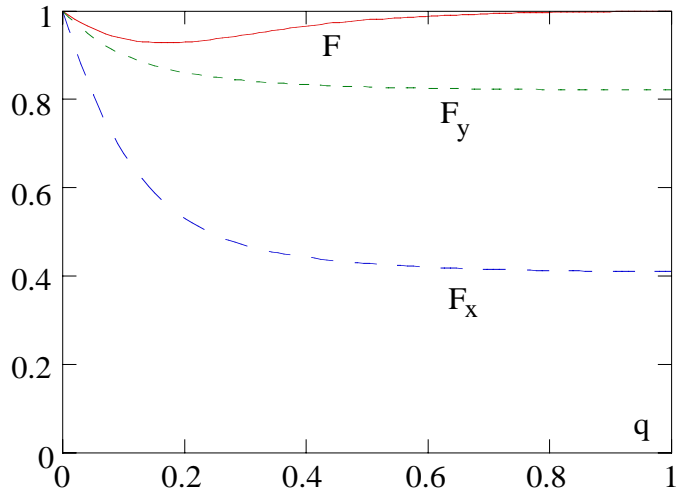

Fig. 17 - $F(u_o)$, and $F_{x,y}(u_o)$ as function of $\bar{q}$

## 7 NON UNIFORM BOUNDARIES

### 7.1 General properties

The uniform boundary cases described in the previous section allow to estimate the effect of smooth pieces of the vacuum chamber on the beam dynamics. Usually we mainly worry about the resistive wall impedance, which produces a shift of the transverse tunes, drives the head-tail and multibunch instabilities. The pipe is, however, interrupted by many devices installed on the machine, RF cavities, diagnostics, wigglers, cross section jumps etc.

Unlike the uniform boundary case, the discontinuities in the vacuum chamber are source of radiated fields which do not travel with the charge. We observe several consequences: excitation of resonant HOMs in resonant structures, new configuration of the self field (after a jump in the cross section), propagation of e.m. fields at frequencies above the cut-off of the beam pipe [18].

As an example, in Fig. 18 we show the case of a relativistic point charge crossing a hole in an infinite p.c. plane.

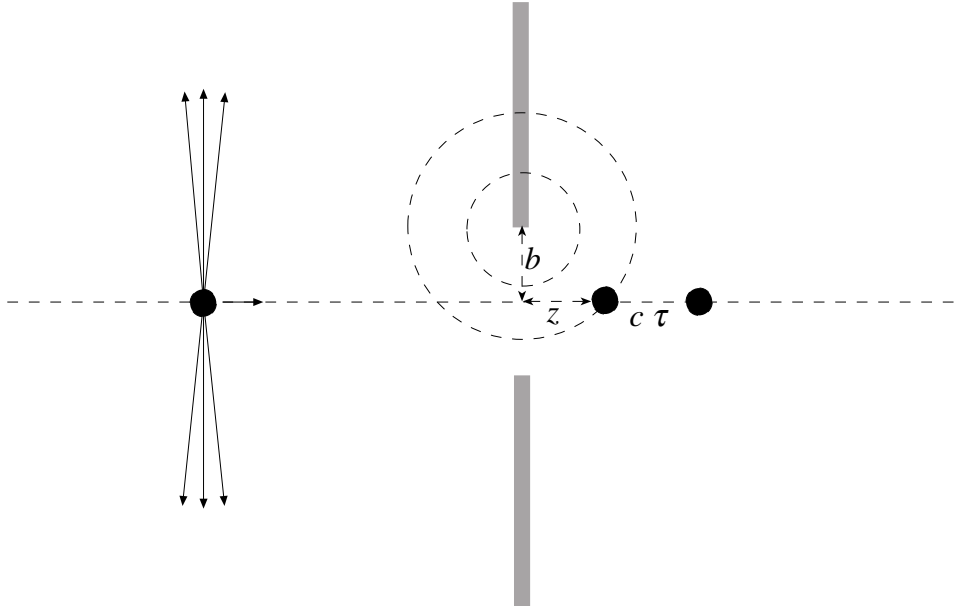

Fig. 18 - Relativistic charge passing through a hole of radius *b*.

The diffraction is caused by the primary fields which, impinging on the hole edge, produce secondary scattered fields propagating at the light velocity. The distance for the radiated fields to catch up the charge itself is $\bar{z} = \gamma b$. A test charge traveling a distance $\beta c \tau$ ($\beta \approx 1$) behind will be reached by the same fields at $\bar{z} \approx (b^2 - c^2\tau^2)/(2c\tau)$.

Another basic feature of the diffraction effects concerns the frequency bandwidth of the power spectrum. Despite the point like nature of the charge, the primary fields exciting the edge have an effective size $\sigma_{eff} = b/\gamma$. The diffraction excitation has a power spectrum extending up to a "radiation cut off frequency" $\omega_{cut-off}^{rad} \approx c\gamma/b$ above which there is an exponential roll off.

The geometry of Fig. 18 has been extensively studied [19]. A ultra relativistic charge passing through the hole loses the energy:

$$U_{11} = 2U(r \geq b) = \frac{3\pi}{8}\left(\frac{r_o}{b}\right)\gamma m_o c^2 \qquad (148)$$

It is interesting to note that the energy loss is twice as much as given by Eq. (111), i.e. the amount of energy stored outside the tube of radius *b*. This feature has been found also for other geometries (such as the step discontinuity) : a ultrarelativistic point charge deposits the same amount of energy in rebuilding the self field as in the radiated fields. This result, in general, can be explained as the typical phenomenon occurring in the charge or discharge of a capacitor.

At low frequencies the longitudinal impedance is [20]:

$$Z(\omega) \approx \frac{Z_o}{2\pi\beta}\left[log\left(\frac{1+\beta}{1-\beta}\right) - \beta + i\pi\right] \qquad (149)$$

## 7.2 A step transition

Let us consider an abrupt change in the cross section of a circular beam pipe from a radius $b$ to radius $d$ (Fig. 19). When the charge crosses the vacuum chamber discontinuity secondary fields are scattered at the sharp edges. The total fields, "primary" plus "secondary" diffracted fields, are such to restore the boundary condition at the pipe walls. This problem has been treated by several authors with numerical and analytical techniques [21,22]. An exact analytical solution has been found for a discontinuity made of two coaxial circular pipes for which both longitudinal and transverse dipole impedances have been derived [23,24]. Here we will report the main relevant results and features.

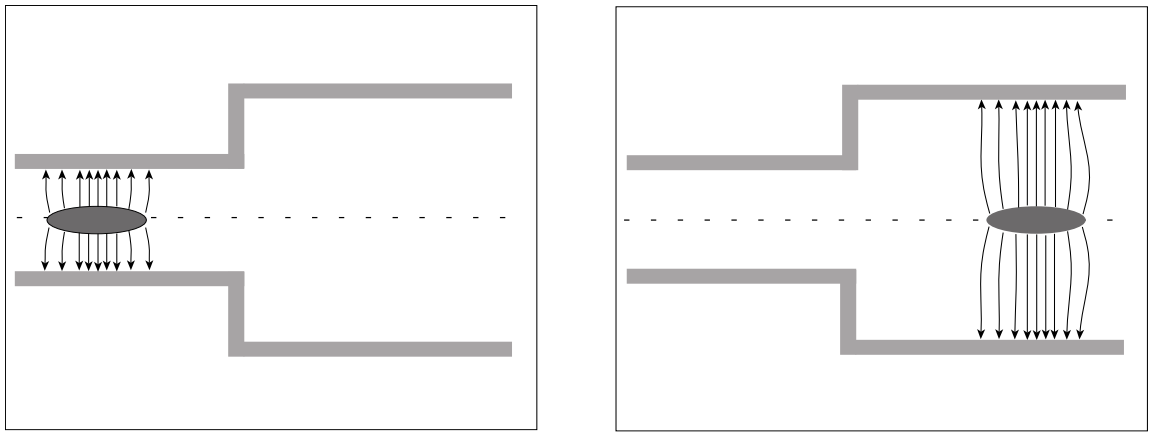

Fig. 19 - Step discontinuity in the beam pipe

We will distinguish two cases: a particle exiting into a beam pipe of a bigger radius, "step-out" case, and a particle entering a narrowing pipe, "step-in" case. Theoretical results show that the impedance is mostly resistive in the step-out case with a big contribution at high frequencies above cut off, while in the step-in case the impedance is low, vanishing at high frequencies. In Fig. 20 we show examples of impedances for the step-in and step-out problems, as derived with the computer code ABCI [25].

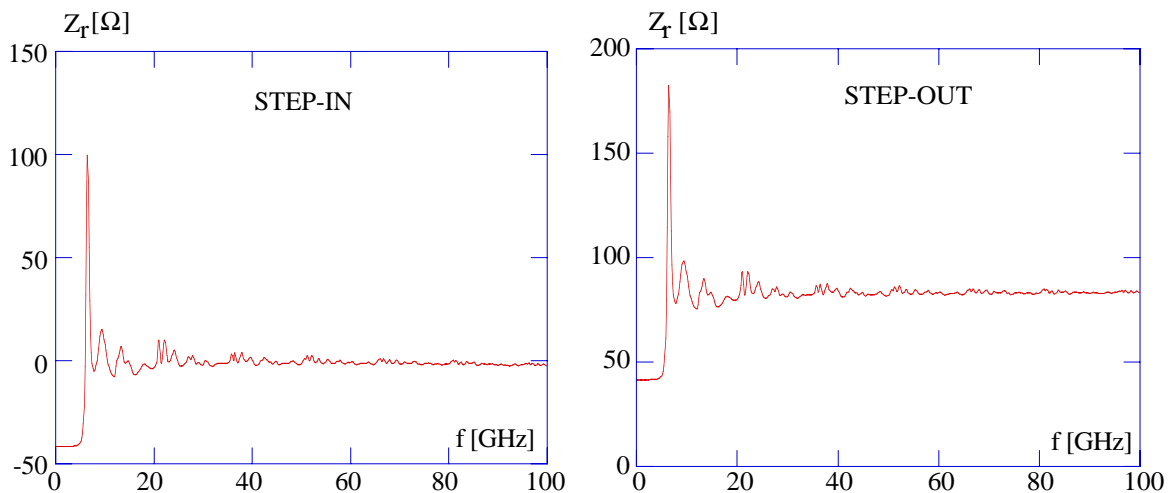

Fig. 20 - Longitudinal impedances of a step-in and step-out discontinuity

The impedances have a resonant behaviour just after the beam pipe cut-off and reaches a constant asymptotic value at high frequencies. The asymptotic behaviour of the real part of the impedance is:

$$Z^{out}_{m=0} \sim \frac{Z_o}{\pi} ln\left(\frac{d}{b}\right) \quad and \quad Z^{in}_{m=0} \sim 0 \tag{150}$$

Such different results are explained by recognizing two main effects contributing to the energy loss. In the step out case, when the charge crosses the discontinuity, the self field restoring the boundary conditions, has to fill the extra space $b < r < d$ between the two pipes, while diffracted fields propagate into the pipes. Both these effects lead to an energy loss that can be put as:

$$q^2 k^{out} = U(b < r < d) + E_{rad} \tag{151}$$

where $E_{rad}$ is the energy radiated at the edges and $U(b < r < d)$ is the energy necessary to fill the region $b < r < d$.

In the step-in case, the radiated energy is reflected back with respect to the particle motion without changing its kinetic energy.

$$q^2 k^{out} = -U(b < r < d) + E_{rad} \tag{152}$$

For a point charge, since the radiated energy is taken out of the energy "missing" in the smaller radius pipe: $E_{rad} \approx U(b < r < d)$, we have:

$$\begin{aligned} q^2 k^{out} &\sim 2U(b < r < d) \\ q^2 k^{in} &\sim 0 \end{aligned} \tag{153}$$

We remind that for a real bunch both $U(b < r < d)$ and $E_{rad}$ depend on the bunch length. In particular, if the bunch spectrum does not cover significantly the frequency region above the pipe cut-off, there is no radiation.

In Fig. 21 we show the dependence of the loss factor on the bunch-length for the step-in and step-out cases, for a step with $b = 2cm$ and $d = 4cm$. In this case we find that a long bunch loses energy in the step-out, but would regains the same amount of energy in a symmetric step-in. Therefore, if a long bunch crosses a pipe enlargement formed by a step-out and step-in sequence with same radii, the total energy loss is almost zero; in this case the wake function inside the bunch is an odd function and the impedance practically inductive.

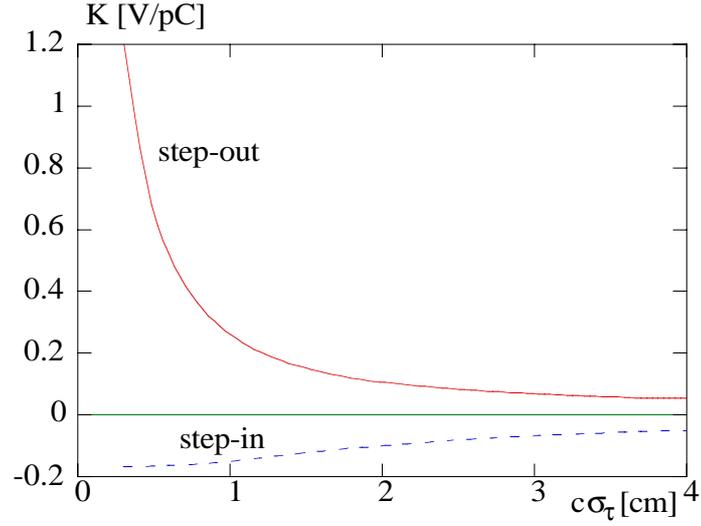

Fig. 21 - Example of step-in and step-out loss factor versus bunch length

The dipole longitudinal and transverse impedance for a discontinuous coaxial pipe has been derived in [24]. The model allows for an exact solution of the e.m. problem, and furnishes simple expressions of the impedance at high frequencies that can be used also for real step transitions:

$$Z^{out}_{m=1} \sim \frac{Z_o}{2\pi^2}\left[\frac{1}{b^2} - \frac{1}{d^2}\right] rr_1 cos(\phi) \quad [\Omega] \quad (154)$$

$$Z^{in}_{m=1} \sim 0$$

The high frequency transverse dipole impedance for the step out case is:

$$\mathbf{Z}'_{\perp,1} \equiv \frac{1}{r_1}\mathbf{Z}_{\perp,1} = \frac{cZ_o}{2\pi^2\omega}\left[\frac{1}{b^2} - \frac{1}{d^2}\right]\hat{r}_1 \quad [\Omega/m] \quad (155)$$

Simple physical approach.

To complete this section we find worth showing that the asymptotic expressions (150) can be derived in with a simple physical approach. We compute the energy $U(b<r<d)$:

$$U(b<r<d) = \varepsilon_o \int_V E_r^2 dV \quad (156)$$

For ultra relativistic charges in a cylindrical pipe, the fields are almost the same as in the free space:

$$E_r \sim \frac{q}{2\pi\varepsilon_o r}\left(\frac{\gamma}{b}\right) \quad (157)$$

Integration (156) gives:

$$U(b<r<d) \sim \frac{q^2 Z_o}{2\pi} \ln\left(\frac{d}{b}\right)\left(\frac{c\gamma}{b}\right)$$

$$k^{out} = 2\frac{\Delta W}{q^2} \sim \frac{Z_o}{\pi} \ln\left(\frac{d}{b}\right)\left(\frac{c\gamma}{b}\right) \tag{158}$$

Remind that $\sigma_{eff} = (b/\gamma)$ is the effective charge size and that as far as $k^{out}$ is inversely proportional to the size, we can expect that $Z_r(\omega)$ is a constant function of frequency (see the discussion in 3.2). Now, applying the definition (72) and considering the spectrum of a bunch with rectangular distribution, we get for a constant impedance:

$$K = \frac{1}{\pi}\int_0^\infty Z_r(\omega) \frac{\sin^2(\omega\sigma_{eff}/2c)}{(\omega\sigma_{eff}/2c)^2} d\omega = \frac{Z_r(\omega)}{\sigma_{eff}/c}. \tag{159}$$

From the comparison (158) and (159) the expected result follows.

### 7.3 Taper

If one uses long gradual tapers instead of the abrupt step transitions the total energy loss may be drastically reduced. Indeed, the infinitely long taper reduces the radiated energy $E_{rad}$ to zero. For a point charge we have:

$$k^{out}_{taper} \sim \frac{U(b<r<d)}{q^2} = \frac{1}{2} k^{out}_{step}$$

$$k^{in}_{taper} \sim -\frac{U(b<r<d)}{q^2} = -\frac{1}{2} k^{out}_{step} \tag{160}$$

It means that in the limit of long tapers the loss factor of a taper-out reaches half the value of the loss factor for a step-out. There may be even an energy gain for the taper-in case. It must be considered, however, that in a vacuum chamber of a circular accelerator there are taper-in and taper-out transitions. As it can be easily seen, long symmetric tapers reduce total losses practically to zero:

$$k = k^{out}_{taper} + k^{in}_{taper} \sim 0 \tag{161}$$

Certainly, we can not use infinitely long tapers in an accelerator design. In [18] it is shown that for a short bunch of rms length $\sigma$ the dependence of the longitudinal loss factor of a one-sided taper on its angle can be approximated by the formula:

$$K = \frac{Z_o c}{2\sigma\pi^{3/2}}\left[1 - \frac{\tilde{\eta}_1}{2}\right]\ln\frac{d}{b} \tag{162}$$

where

$$\tilde{\eta}_1 = min\left\{1, \frac{g\sigma}{(d-b)^2}\right\} \quad (163)$$

For a symmetric taper $\tilde{\eta}_1 / 2$ is replaced by $\tilde{\eta}_1$. So, the condition:

$$\frac{g\sigma}{(d-b)^2} > 1 \quad (164)$$

can be considered as an approximate criterion to choose a reasonable taper length. We should say here that the formula (162) is valid for short bunches when the main contribution to the losses comes from the high frequency impedance and the diffraction model [18,26] can be applied. Because of that we advice the reader to use numerical codes in order to check the criterion (164) for any particular case. As an example, we show in Fig. 22 the loss factor versus the taper length $g$ for a tapered out structure passing from $b = 2$ cm to $d = 4$ cm. One can see that the loss factors for a bunch of 0.5 cm and 1.0 cm computed by the code ABCI, approach an asymptotic value corresponding to the case of no radiation, at a value of $g$ such that $\tilde{\eta}_1 = 1$. For the 3 cm bunch length the whole bunch spectrum lies below the beam pipe cut-off and no radiation occurs.

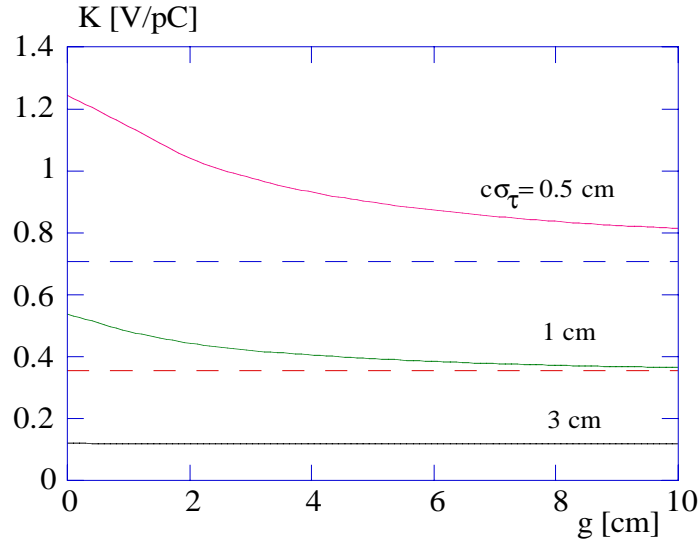

Fig. 22 - Loss factor of a tapered discontinuity

## 7.4 Single cell cavity

Cross section variations in an accelerator vacuum chamber can create resonant cavities. Part of the fields excited in the cavities is entrapped reflecting back and forth generating the resonant modes. Above cut-off the amplitudes of the resonance drops because of the energy leakage into the vacuum chamber, the resonances overlap leading to the smooth, "broadband" impedance.

A typical cavity impedance is shown in Fig. 23 At the frequencies below $\omega_c$ a real high-Q cavity has many sharp resonance. In a RF cavity the fundamental one is used to supply energy to the beam; all the others are "parasitic" modes (higher order modes - HOM) which subtract energy from the beam. Above cut-off the resonances are broadened.

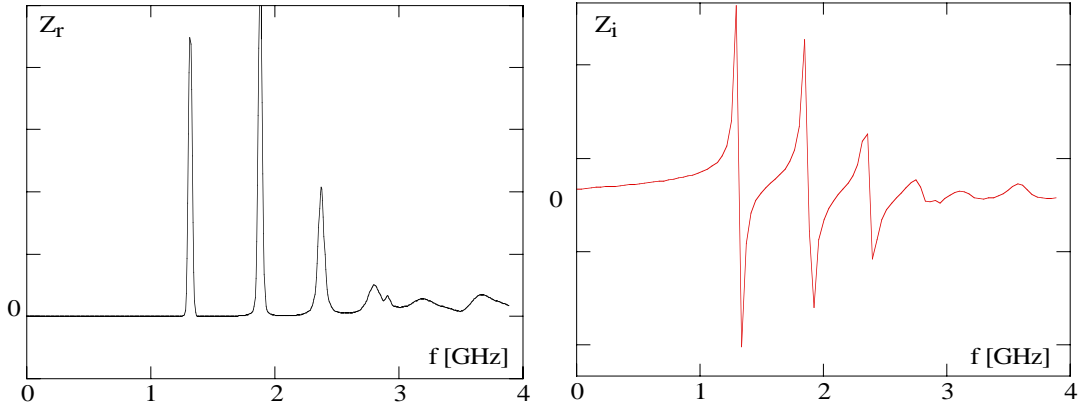

Fig. 23 - Typical impedance spectrum for a cavity with attached tubes

### 7.4.1 Monopole HOM (longitudinal)

In Sec. 2.2.1 we have found the wake potential of a single HOM. Following the results (47) of Sec. 2.8 for cylindrical symmetry, we can write the longitudinal wake of a monopole HOM (m=0) as:

$$w_{z,o}(\boldsymbol{r},\boldsymbol{r}_1;\tau) = 2k_o(r,r_1)e^{-\Gamma_o\tau}\left[\cos(\overline{\omega}_o\tau) - \frac{\Gamma_o}{\overline{\omega}_o}\sin(\overline{\omega}_o\tau)\right]H(\tau) \qquad (165)$$

$$\text{with} \quad \Gamma_o = \frac{\omega_o}{2Q_o} \ , \ \overline{\omega}_o^2 = \omega_o^2 - \Gamma_o^2 \qquad (166)$$

In Sec. 2.9 we have also found that in the ultra relativistic limit the monopole longitudinal wake does not depend on the radial displacement of both leading and trailing charges (54). This result is conveniently exploited in the numerical codes where the loss factor $k_o(r,r_1)$ is computed at the pipe radius, thus limiting the calculation of the energy loss over a definite and limited path:

$$k_o(r,r_1) \equiv k_o(b) = \frac{\omega_o R_o}{2Q_o} = \frac{|V_o(b)|^2}{2U_o} \qquad (167)$$

where $V_o(b)$ is the voltage gain computed at $r = b$ and $U_o$ is the average energy stored in the HOM.

Applying the Fourier transform to (165) we get the longitudinal impedance of a monopole resonant HOM:

$$Z(\omega) = \frac{R_o}{1 + jQ_o\left(\dfrac{\omega}{\omega_o} - \dfrac{\omega_o}{\omega}\right)} \tag{168}$$

It is interesting to note that the shunt impedance is also defined as:

$$R_o = \frac{|V_o(b)|^2}{P_{od}} T^2 \tag{169}$$

where $P_{od}$ is the power dissipated at the cavity wall or in any damping device (loops, waveguides etc.), and $T$ is the transit time factor defined as the ratio between the accelerating voltage seen by a traveling charge and the voltage at the gap:

$$T = \frac{1}{\displaystyle\int_{gap} E_z\, dz} \left|\int_{gap} E_z\, e^{jkz} dz\right| \tag{170}$$

which takes into account the time evolution of the fields during the cavity crossing. The transit time factor approaches unity at low frequencies (wavelength much bigger than the gap).

In the low frequency limit $\omega \to 0$ the impedance is purely inductive. In case of n HOMs we have:

$$Z(\omega) = j\omega \sum_n \left[\frac{R_n}{Q_n \omega_n}\right] = j\omega L \tag{171}$$

### 7.4.2 Dipole HOM (longitudinal)

The wake potential of a single dipole (m=1) HOM for cylindrical symmetry is given by:

$$w_{z,1}(\mathbf{r},\mathbf{r}_1;\tau) = 2\cos(\phi)k_1(r,r_1)e^{-\Gamma_1 \tau}\left[\cos(\overline{\omega}_1 \tau) - \frac{\Gamma_1}{\overline{\omega}_1}\sin(\overline{\omega}_1 \tau)\right]H(\tau) \tag{172}$$

$$\text{with}\quad \Gamma_1 = \frac{\omega_1}{2Q_1}\ ,\ \overline{\omega}_1^2 = \omega_1^2 - \Gamma_1^2 \tag{173}$$

The dipole wake potential in the ultra relativistic limit is proportional to the transverse displacements of both charges. Therefore, we may scale the loss factor computed at the pipe radius as:

$$k_1(r, r_1) \equiv k_1(b) \frac{rr_1}{b^2} \tag{174}$$

$$k_1(b) = \frac{\omega_1 R_1}{2Q_1} = \frac{|V_1(b)|^2}{2U_1} \tag{175}$$

$$Z_1(r, r_1; \omega) = \frac{R_1}{1 + jQ_1\left(\frac{\omega}{\omega_1} - \frac{\omega_1}{\omega}\right)} \frac{rr_1}{b^2} \cos(\phi) \tag{176}$$

### 7.4.3 Dipole HOM (transverse)

The general relationship between transverse and longitudinal wake functions allows to obtain the transverse dipole wake function (96):

$$\mathbf{w}_{\perp,1}(\mathbf{r}, \mathbf{r}_1; \tau) = \mathbf{r}_1 \frac{2c}{\overline{\omega}_1 b^2} k_1(b) e^{-\Gamma_1 \tau} \sin(\overline{\omega}_1 \tau) H(\tau) \tag{177}$$

The transverse impedance (102) is:

$$\mathbf{Z}_{\perp,1}(\mathbf{r}, \mathbf{r}_1; \omega) = \left(\frac{c}{\omega}\right) \frac{\frac{R_1}{b^2}}{1 + jQ_1\left(\frac{\omega}{\omega_1} - \frac{\omega_1}{\omega}\right)} \mathbf{r}_1 \tag{178}$$

sometimes expressed as:

$$\mathbf{Z}_{\perp,1}(\mathbf{r}, \mathbf{r}_1; \omega) = \left(\frac{\overline{\omega}_1}{\omega}\right) \frac{R'_{\perp,1}}{1 + jQ_1\left(\frac{\omega}{\omega_1} - \frac{\omega_1}{\omega}\right)} \mathbf{r}_1 \tag{179}$$

with

$$R'_{\perp,1} = \frac{cR_1}{\overline{\omega}_1 b^2} = \frac{\overline{\omega}_1}{c} R_{\perp}^{Urmel} \tag{180}$$

The quantities $R_1$ and $R'_{\perp,1}$ are derivable from the output data of numerical codes like URMEL.

### 7.4.2 High frequency case

The high frequency impedance is mainly due to the interaction of the charges with the fields diffracted at the cavity-pipe edge. When the bunch length is much smaller than the beam pipe radius, the high frequency contribution to energy loss and impedance can be dominant. Numerical calculations require significant time to find the wake potentials of very short bunches, therefore an analytical analysis of the high frequency behaviour of the impedance (short range wake potential) becomes necessary. Methods of diffraction theory are used to calculate the impedance at the high frequencies, $\omega \gg c/b$. For a pill-box cavity of length $g$ radius $d$, Fig. 24, and with side pipes of radius $b$ the diffraction model gives [27,28,29]:

$$Z(\omega) = (1-j)\frac{Z_o}{2\pi b}\sqrt{\frac{gc}{\omega\pi}} \qquad (181)$$

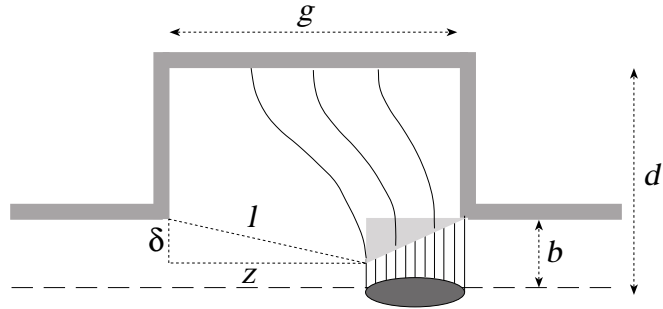

Fig. 24 - Relevant parameters for a pill box with attached tubes.

The formula is valid for the region of parameters where $g \ll kb^2$ (a 'cavity regime'). For the region of parameters $g \gg kd^2$ the diffraction model of [26] gives the same impedance as found for a "step":

$$Z(\omega) = \frac{Z_o}{\pi}\ln\frac{d}{b} \qquad (182)$$

In the transition region of parameters $kb^2 \ll g \ll kd^2$

$$Z(\omega) = \frac{Z_o}{2\pi}\ln\frac{gc}{\omega b^2} \qquad (183)$$

The transition from the cavity to the step regime is explained by the transition from the Fresnel diffraction for the cavity to the Fraunhofer diffraction for the step. It was shown that the transition from one regime to another occurs when:

$$\eta = \frac{2g\sigma}{(d-b)^2} \sim 1 \qquad (184)$$

The cavity regime is reached when the parameter $\eta \ll 1$. In the opposite case, $\eta \gg 1$, the regime of a step is fulfilled.

For a short Gaussian bunch, for which $\sigma \ll b$, the high frequency tail of the impedance mainly contributes to the energy loss. In this case the longitudinal and transverse loss factors are given by:

$$K = \frac{Z_o c}{4\pi^2 b} \Gamma\left(\frac{1}{4}\right) \sqrt{\frac{g}{\pi\sigma}}, \quad \Gamma\left(\frac{1}{4}\right) = 3.6256..$$

$$K_\perp = \frac{Z_o c}{4\pi} \frac{\sqrt{\pi g \sigma}}{b^3} \qquad (185)$$

Simple physical approach

The high frequency behaviour of the impedance and loss factor is qualitatively explained by the features of the fields diffracted at the cavity edges [30]. Consider a bunch of charge $q$ and rms size $\sigma$ passing a pill-box cavity (Fig. 24). The field excited by the head of the bunch and diffracted by the left edge, touches the bunch tail if:

$$\frac{l}{c} = \frac{z+\sigma}{\beta c} \quad with \quad l = \sqrt{\delta^2 + z^2} \qquad (186)$$

When the bunch leaves the cavity, for z=g, the self-field has been perturbed in the shadow region with $\delta \sim \sqrt{2\sigma g}$. The bunch has to rebuild the field in this region in order to restore the boundary condition in the rhs beam pipe. The e.m. energy removed in the shadowed region of Fig. 24 is:

$$U \sim \frac{q^2}{4\pi\varepsilon_o \sigma} \int_{b-\delta}^{b} \frac{dr}{r} \qquad (187)$$

if $\delta \ll b$:

$$U \sim \frac{q^2}{4\pi\varepsilon_o \sigma} \frac{\delta}{b} = \frac{q^2}{4\pi\varepsilon_o \sigma} \frac{\sqrt{2\sigma g}}{b} \qquad (188)$$

thus giving:

$$K = \frac{U}{q^2} = \frac{1}{4\pi\varepsilon_o b} \sqrt{\frac{2g}{\sigma}} \qquad (189)$$

Note that the condition $\delta \ll b$ implies also that $\sigma \ll b$ (g is comparable with b) and most part of the bunch spectrum lies well above the beam pipe cut-off frequency. As far as the loss factor is inversely proportional to the bunch length the impedance scales as $1/\sqrt{\omega}$. For a point charge the effective bunch length is b / $\gamma$ giving the scaling for the loss factor $k \sim \sqrt{\gamma}$.

### 7.5 Periodic r.f. structure

An array of periodic cavities shows an impedance spectrum similar to the single cell cavity below the cut-off frequencies. Many sharp resonant modes appear in the impedance spectrum corresponding to the normal modes of the single cavities but also to the coupled modes between cells, like in a system of coupled oscillators. Above cut-off, however, the broad impedance decays at high frequencies with the asymptotic law $\omega^{-3/2}$ [31,32]. A qualitative picture of the diffraction phenomenon at the basis of such behaviour is given in [30]. Basically, the energy refilling becomes much less in a periodic structure, since the self-field has no time to be rebuild from one cell to the next. The transition from the single cell regime to the periodic structure regime has also been studied [33]. For a structure of $M$ rf cells of period $d$ and iris radius $b$, we have the transition from the single cell regime to the periodic structure regime at the frequency:

$$\omega_t \approx \frac{cMd}{b^2} \qquad (190)$$

above which the impedance follows the single cell asymptotic behaviour $\omega^{-1/2}$. It is worth noting that in the periodic structure, the asymptotic decay $\omega^{-3/2}$ makes the loss factor independent from the bunch length and from the energy. Similar expressions have been found for the case of a charge passing through an array of holes in infinite planes, reproducing the same diffraction phenomenon [34,35].

### 7.6 Small discontinuities

In a real vacuum chamber there are many small discontinuities such as shallow cavities, tapers, masks, bellows, etc. In spite of the little size, their overall contribution to the inductive impedance, up to rather high frequencies, cannot be neglected. These elements give the main contribution to the longitudinal inductive impedance responsible for the potential-well bunch lengthening process.

At low frequencies a satisfactory estimate of the inductive impedance $Z(\omega) = j\omega L$ can be obtained by applying Faraday's law in the static approximation. However, it has been shown that one can get more accurate results taking into account a correction factor coming from distortion of the electric field at the chamber discontinuity [36]. Here we do not discuss the method used by authors, but reproduce helpful expression for the impedance of some typical discontinuities shown in Figs. 25.

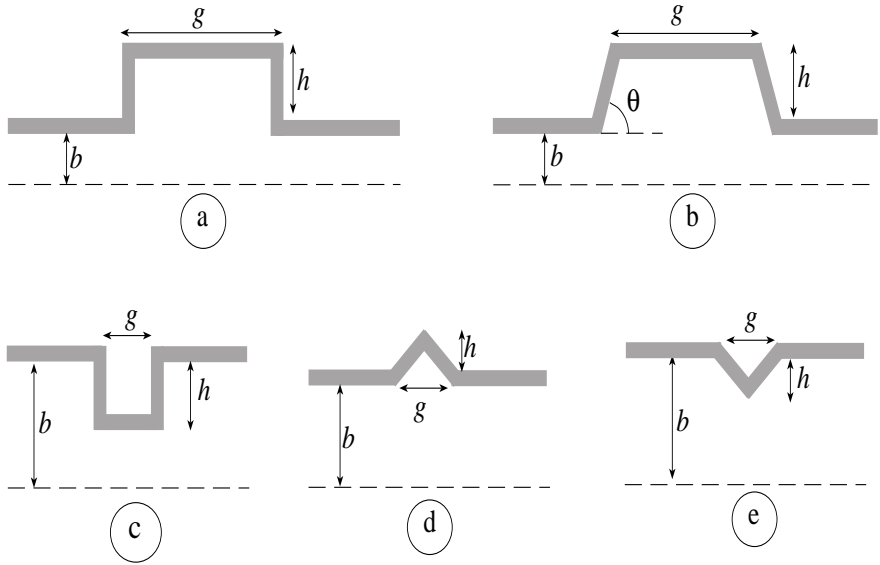

Fig. 25 - Examples of vacuum chamber discontinuities

### 7.5.1 Shallow cavity (Fig. 25a)

The low frequency impedance of the small short pill-box ($g < h$) is given by:

$$Z(\omega) = j\omega \frac{Z_o}{2\pi bc}\left(gh - \frac{g^2}{2\pi}\right) \tag{191}$$

For the opposite extreme, $g >> h$, but still $g < b$:

$$Z(\omega) = j\omega \frac{Z_o h^2}{2\pi^2 bc}\left(2ln\left(\frac{2\pi g}{h}\right) + 1\right) \tag{192}$$

When the length $g$ of the pill-box is greater than the pipe radius $b$, the pill-box can be considered as composed of two independent steps each giving the following contribution to the impedance:

$$Z(\omega) = j\omega \frac{Z_o h^2}{4\pi^2 bc}\left(2ln\left(\frac{2\pi b}{h}\right) + 1\right) \tag{193}$$

The transverse impedance of the shallow cavity, if the assumption $(g,h) < b$ is valid, is given by:

$$Z_\perp = jZ_o \frac{g}{\pi b^2}\left(\frac{d^2 - b^2}{d^2 + b^2}\right) \tag{194}$$

where $d = b + h$.

### 7.5.2 Shallow trapezoid (Fig. 25b)

For a long shallow trapezoid ($g >> h$) the impedance of a single sloping step (taper) with the slope angle $\theta = \pi v$ has the form:

$$Z(\omega) = j\omega \frac{Z_o h^2}{2\pi^2 bc} \left[ \ln \pi v \left( \frac{b}{h} - 2\cot \pi v \right) + \frac{3}{2} - \gamma - \psi(v) - \frac{\pi}{2} \cot \pi v - \frac{1}{2v} \right] \quad (195)$$

where $\gamma = 0.5772...$ is Euler's constant, $\psi(v)$ is the "psi" function and the transition is assumed to be short compared to the chamber radius, i. e. its length $l = h \cot \pi v << b$.

### 7.5.3 Shallow iris (Fig. 25c)

Two extreme cases can be considered for the iris geometry. When $g >> h$ the impedance of the iris coincides with that of a shallow cavity. For the case of a thin (or deep) iris $g << h$ the expression for the longitudinal impedance takes the form:

$$Z(\omega) = j\omega \frac{Z_o}{4bc} \left\{ h^2 + \frac{gh}{\pi} \left[ 2\ln(8\pi g / h) - 3 \right] \right\} \quad (196)$$

### 6.5.4 Discontinuities of a triangular shape (Fig. 25d,e)

A short discontinuity of a triangular-shaped cross-section with height $h$ and base $g$ ($g << h$) has the following low-frequency impedance:

$$Z(\omega) = j\omega \frac{Z_o}{4\pi bc} \left( gh - \frac{g^2}{\pi} \right) \quad (197)$$

The impedance of a triangular iris is given by:

$$Z(\omega) = j\omega \frac{Z_o}{4bc} \left[ h^2 + \frac{2gh}{\pi} (1 - \ln 2) \right] \quad (198)$$

For the case of shallow triangular perturbations, $h << g < b$, both the enlargement and iris have the same inductive impedance, independent of $g$:

$$Z(\omega) = j\omega \frac{2Z_o h^2 \ln 2}{\pi^2 bc} \quad (199)$$

## 7.7 Elements of Beam Diagnostics

Each accelerator has numerous diagnostics elements such as button pick-ups, strip-line beam position monitors, etc.. Impedance calculation of such elements is a rather difficult task. Being inserted into a vacuum chamber a diagnostics element brakes the vacuum chamber symmetry and one has to analyze the interaction of a bunch with a complicated 3-dimensional structure.

Numerical solution to the problem is also not a simple task. Usually, sizes of the diagnostics element are small. This demands a very fine mesh, and so large computation time. Moreover, the image currents induced in the elements flow into the external circuits. The correct simulation of this external load appears to need additional analytical efforts or using recently developed sophisticated numerical codes.

The methods which allows an analytical treatment of the impedance at low frequencies, $\omega << \omega_c$, is based on the electrical approach:
- the 3D geometry is substituted by an equivalent circuit consisting of concentrated radio technical elements and transmission lines;
- methods of electric circuits and theory of transmission lines are used to find currents and voltages in the circuit elements;
- in the case of matched load, the coupling impedance is defined by considering the power lost by a bunch to be equal to that dissipated in the load. Some other consideration can be taken into account to relate the beam coupling impedance with the currents and voltages in the circuit elements.

Here we will closely follow the treatment of [39] in order to illustrate the method. Let us consider a strip-line pick-up of the length l covering the azimuthal angle $\phi$ (see Fig. 26).

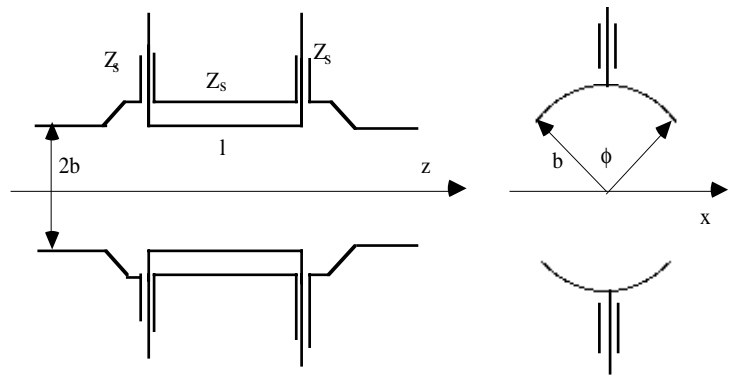

Fig. 26 - Strip line pick-up

The strip plate together with the vacuum chamber wall create a transmission line of the characteristic impedance $Z_s$. If the transmission line is terminated by a matching resistance $Z_s$ at the each end, one can draw the equivalent scheme of Fig. 27.

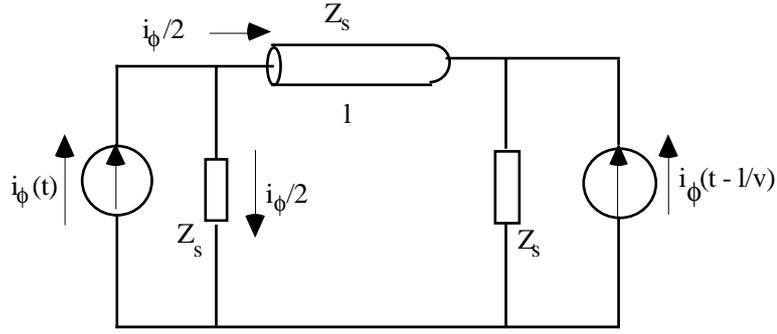

Fig. 27 - Equivalent circuit for a strip line

where $i_\phi(t)$ is the fraction of the image current intercepted by the strip plate:

$$i_\phi(t) = -\left(\frac{\phi}{2\pi}\right) i_b(t)$$

with $i_b(t)$ the beam current.

In the absence of a dielectric the signal velocity in the transmission line is equal to that of light. Moreover, as it can be seen from the equivalent scheme, the strip-line monitor possesses directional property for the relativistic particle with β~c. In fact the signal appears only across the upstream port (with respect to the beam velocity):

$$V(t) = \frac{Z_s}{2}\left(\frac{\phi}{2\pi}\right)\left[i_b(t) - i_b\left(t - \frac{2l}{c}\right)\right] \tag{200}$$

In turn, only a fraction $\phi/2\pi$ of the total image current "sees" the potential difference, from what follows that the average potential seen by a beam is:

$$V_b(t) = V(t)\left(\frac{\phi}{2\pi}\right) \tag{201}$$

Applying the definition of the impedance as a response for the sinusoidal current perturbation, we have:

$$Z(\omega) = Z_s\left(\frac{\phi}{2\pi}\right)^2\left[sin^2(\kappa l) + j\,sin(\kappa l)cos(\kappa l)\right] \tag{202}$$

The same result for the real impedance has been derived by considering the real power dissipated in the upstream termination as the power lost by a bunch [38]. The imaginary part was found by Hilbert transform.

The transverse impedance of a pair of symmetric strip-lines matched on the both ends has been found in [39]. In the direction perpendicular to the strip-lines:

$$Z_\perp(\omega) = \frac{c}{b^2}\left(\frac{4}{\phi}\right)^2 \left(sin^2 \frac{\phi}{2}\right)\left[\frac{Z(\omega)}{\omega}\right] \quad (203)$$

where Z(ω) is the longitudinal impedance of that pair of striplines. A particle shifted with respect to the system axis in the direction parallel to the strips does not experience any transverse kick and the transverse impedance in the direction $Z_\perp(\omega) = 0$.

The same method gives for a strip-line forming a transmission line with the beam pipe with characteristic impedance $Z_s$ and terminated at the center by $Z_s$ the following longitudinal impedance [39]:

$$Z(\omega) = Z_s \left(\frac{\phi}{2\pi}\right)^2 (1 - j\kappa l)(1 - e^{-j\kappa l}) \quad (204)$$

where $l$ is again the strip length and $\phi$ is the azimuthal angle characterizing the fraction of the image current intercepted by the strip plate.

The longitudinal coupling impedance of a small button-like pickup of a radius r terminated trough a coaxial cable of the characteristic impedance $Z_s$ by a resistance R = $Z_s$ is [40]:

$$Z(\omega) = \left(\frac{\omega_1}{\omega_2}\right)^2 \frac{\omega/\omega_1}{[1+(\omega/\omega_1)^2]}\left[\frac{\omega}{\omega_1} + j\right] \quad (205)$$

with

$$\omega_1 = \frac{1}{RC} \quad \text{and} \quad \omega_2 = \frac{2bc}{r^2}$$

where b is the beam pipe radius, C is the capacitance between the button and the beam pipe wall which is roughly estimated as: $C \sim \varepsilon_0 \pi r^2 / \delta$ for $\delta << b$ being the gap between the wall and the button.

Many examples can be found in the literature on the subject. However, we should say that the high frequency behaviour of the diagnostics elements has not been satisfactory investigated yet. One of the major problems, for the further investigation are the high frequency resonances that can be excited in the structures formed by a diagnostics element and beam pipe walls. Some of such resonances can be associated with standing waves, which do not dissipate their power in the external terminations [41]. Obviously, the high frequency resonances will give some additional inductive contribution to the low frequency impedance.

## 7.8 Holes and Slots in the vacuum chamber.

In order to reduce the coupling impedance due to pumping volumes shielding screens with a number of holes or slots are used. The number of such holes and slots may be large in an accelerator and their contribution to the impedance have to be estimated. It is obvious that because of absence of the axial symmetry a numerical solution of the problem is essentially three dimensional. This implies very time-consuming computations even in the case of a simplified model.

The method which allows analytical calculation of the impedance at low frequencies is based on the Bethe theory of diffraction by a small hole [42]. According to the theory, the small hole is excited by the incident electromagnetic waves created by a given current perturbation. Then, the diffracted fields can be obtained replacing the hole by effective surface "magnetic" currents, which are necessary to satisfy the boundary conditions on the hole. The coupling impedance is found by integrating the fields along the beam trajectory. At low frequencies, $\omega \ll c/b$, in the case of a small hole of radius $h \ll b$ the impedance can be calculated in terms of hole polarizability [43,44]. For a circular hole it gives:

$$Z(\omega) = j\omega \frac{Z_o}{6\pi^2} \frac{h^3}{cb^2} \qquad (206)$$

As far as a single hole introduces the axial asymmetry, the transverse impedance depends on the angle between the beam-offset and the direction to the hole:

$$\boldsymbol{Z}_\perp(\omega) = jZ_o \frac{2h^3}{3\pi^2 b^4} \boldsymbol{a}_r \cos\theta \qquad (207)$$

where $\boldsymbol{a}_r$ is the unit vector to the hole and $\theta$ is the azimuthal angle between the direction and the beam-offset.

It is worth noting, that for the number of holes $M \geq 3$ uniformly spaced in one cross section the restoration of the axial symmetry occurs in a sense that the transverse kick is in the direction of the beam transverse displacement and the resulting impedance does not depend on the azimuthal positions of the holes:

$$\boldsymbol{Z}_\perp(\omega) = jZ_o \frac{h^3}{3\pi^2 b^4} M \boldsymbol{r}_1 \qquad (208)$$

with $\boldsymbol{r}_1$ being the unit vector in the direction of the beam displacement. The coupling impedance of the circular hole falls with wall thickness t, reaching 56% of the value (206) for $t/h > 2$ [44].

The real part of the hole impedance, responsible for the energy losses, is usually neglected because it is much smaller than the imaginary impedance. It is given by [43]:

$$Z_r(\omega) = \frac{5Z_o}{54\pi^3}\left(\frac{\omega h}{c}\right)^4 \frac{h^2}{b^2} \tag{209}$$

The same method gives simple analytical expressions for the impedance of small longitudinal elongated elliptical slot of width w and length l such as $w \ll l \ll b$:

$$Z(\omega) = j\frac{Z_o}{96\pi}\frac{\omega}{c}\frac{w^4}{b^2 l}\left(\ln\frac{4l}{w} - 1\right) \tag{210}$$

$$\mathbf{Z}_\perp(\omega) = j\frac{Z_o}{24\pi}\frac{w^4}{b^4 l}\left(\ln\frac{4l}{w} - 1\right)\mathbf{a}_r \tag{211}$$

Analytical results are available also for the small transverse narrow elliptical slot with $w \ll l \ll b$, cutting the wall image currents [45]:

$$Z(\omega) = j\frac{Z_o}{96\pi}\frac{\omega}{c}\frac{l^3}{b^2\left(\ln\left(\frac{4l}{w}\right) - 1\right)} \tag{212}$$

$$\mathbf{Z}_\perp(\omega) = jZ_o\frac{l^3}{24\pi b^4\left(\ln\left(\frac{4l}{w}\right) - 1\right)}\mathbf{a}_r \cos\theta \tag{213}$$

More discussion on the small slot impedance calculations and estimates can be found in [46].

When the sizes of slots and holes are comparable or larger than the beam pipe radius the static solution to the problem is no longer valid. In that case analytical solutions are known for two particular cases of a long narrow slot and a narrow gap.

It is clear from the physical point of view that the narrow ($w \ll b$) long ($l \gg b$) slot distorts the fields inside a vacuum chamber only by the slot ends. Because of that the impedance should not depend on the length of such a slot. Indeed, according to the perturbation method [47] the impedance of a narrow long (but finite) rectangular slot is a constant for $\omega l / c \gg 1$:

$$Z(\omega) = j\frac{Z_o \pi w^2}{4(2\pi)^3 b^2} \tag{214}$$

The impedance for the case of narrow transverse gap of width g << b separating two pieces of the infinite beam pipe has been derived in Ref.[48]. In the case $\omega g / c << 1$ the longitudinal impedance of the gap is capacitive:

$$Z(\omega) = -j \frac{Z_o c}{2\omega b \ln(16b/g)} \tag{215}$$

We should note here that for a beam pipe of a general cross-section in order to reduce the coupling impedance it is preferable to distribute holes in places where fields created by a bunch are minimal, for example, in corners [45]. In particular, for a rectangular beam pipe of width a and height b the longitudinal impedance of a hole situated on the lateral side ($x_h = \pm a/2$) and displaced by $y_h$ from the horizontal plane of symmetry ($|y_h| < b/2$) is proportional to:

$$Z(\omega) \sim \left( \sum_{m=0}^{\infty} \frac{\cos(2m+1)\pi y_h / b}{\cosh(2m+1)\pi x_h / b} \right)^2 \tag{216}$$

As it can be easily seen, the closer to the corner ($|y_h| \to b/2$) the smaller impedance we have. This is certainly valid for small holes.

Discussions on possible coherent effects of many holes can be found in [49,50]. We just mention that the coherence can be destroyed using non-uniform hole spacing.

## 8. BROAD BAND IMPEDANCE MODELS

The impedance of an accelerator is usually a very complicated function of frequency with many sharp peaks. This complexity of the impedance makes an analytical treatment almost impossible. However, in the study of the single bunch dynamics, one finds that the wake potentials over the bunch length is of main interest. This implies that, in the frequency domain, the bunch can not resolve the details of the sharp resonances and it rather experiences an average effect (the peaks are smeared out). In order to demonstrate this we calculate the wake potentials created by a bunch passing through a cavity, by means of the ABCI code.

The longitudinal impedance is found by Fourier transform of the wake potentials.

As it is clearly seen in Figs. 28 and 29, if one is interested in shorter range part of the wake the resonant impedance peaks are transformed into the smoother and broader impedance.

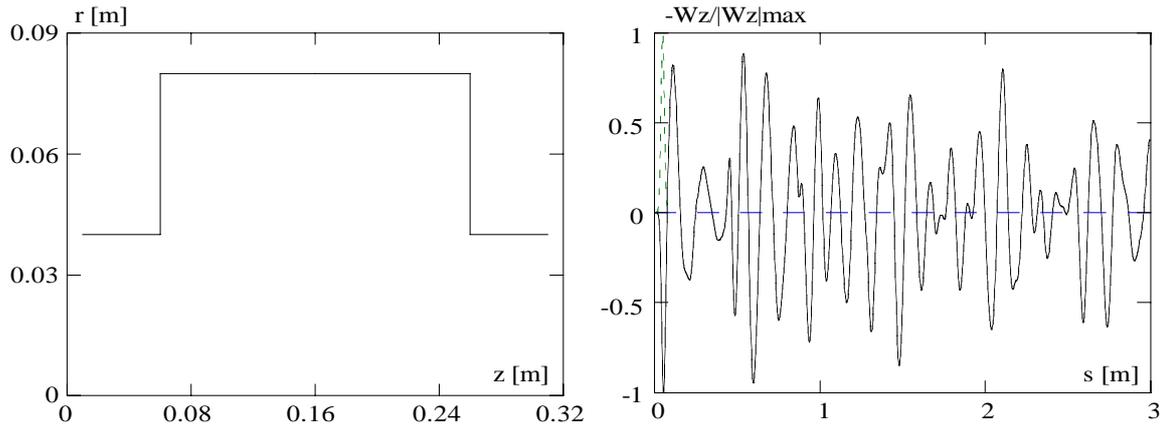

Fig. 28 - Pill-box cavity with tubes and wake function for $c\sigma_\tau = 3$ cm

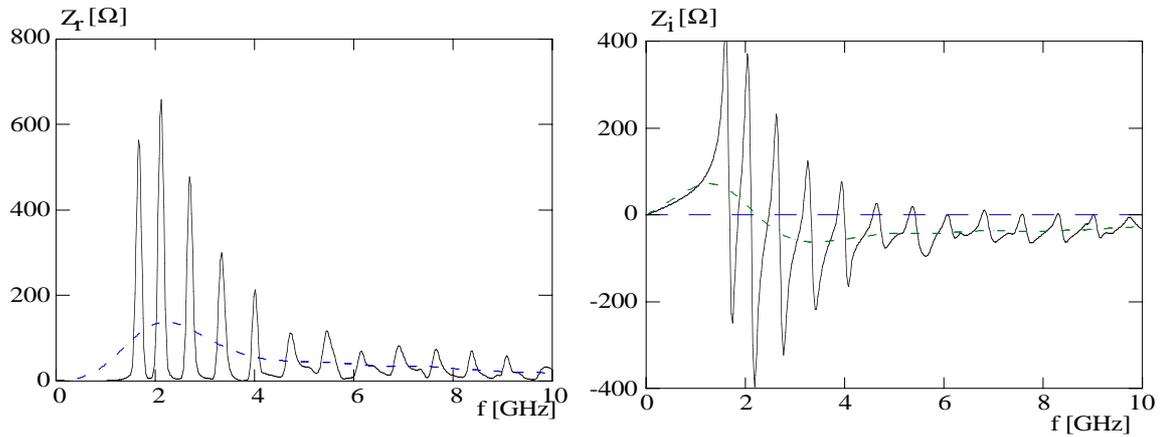

Fig. 29 - Broad Band Impedance: FFT of the wake function over 3 m (solid line);
FFT of the wake function over 30 cm (dashed line).

Therefore, the actual impedance can be replaced by some "broad-band model impedance", which usually is characterized by a small number of parameters, allowing analytical evaluation of stability limits or growth rates of single bunch instabilities.

### 8.1 Broad-band resonator model

The reduced frequency resolution has been the main justification for using the broad-band resonator model [51], historically being introduced first.

The actual impedance is replaced by that of a single mode cavity with a low quality factor $Q$:

$$Z(\omega) = \frac{R_s}{1 + jQ\left(\dfrac{\omega}{\omega_r} - \dfrac{\omega_r}{\omega}\right)} \tag{217}$$

So, only 3 parameters, the shunt impedance $R_s$, angular resonant frequency $\omega_r$, quality factor $Q$, are necessary to describe the impedance frequency behaviour.

For long bunches a cavity shunt impedance $R_s$ is estimated by averaging the resistive part of parasitic resonance (which can be measured by means of perturbation method, or estimated by computer codes) to give the same energy loss as the whole cavity. Each resonance contributes with only half the area of its spectrum.

The quality factor is usually taken as $Q = 1$, while the resonant frequency $\omega_r = \omega_{\text{cutoff}} = 2.4\,(c/b)$, i. e. frequency cut-off of the cavity iris of the radius $b$. These choices are somewhat arbitrary and are satisfactory only for the case of long bunches, when the whole bunch spectrum lies within the beam-pipe cut-off. As an example, let us consider again the cavity of Fig. 28. We find out that the effective impedance calculated as Fourier transform of the wake over 30 cm distance behind the bunch head is approximated reasonably well by the broad band resonator impedance with $R_s = 138\ \Omega$, $Q = 1$ and $f_r = 2.2$ GHz (which is close to the $f_{\text{cut-off}} = 2.5$ GHz of the considered cavity) (Fig. 30).

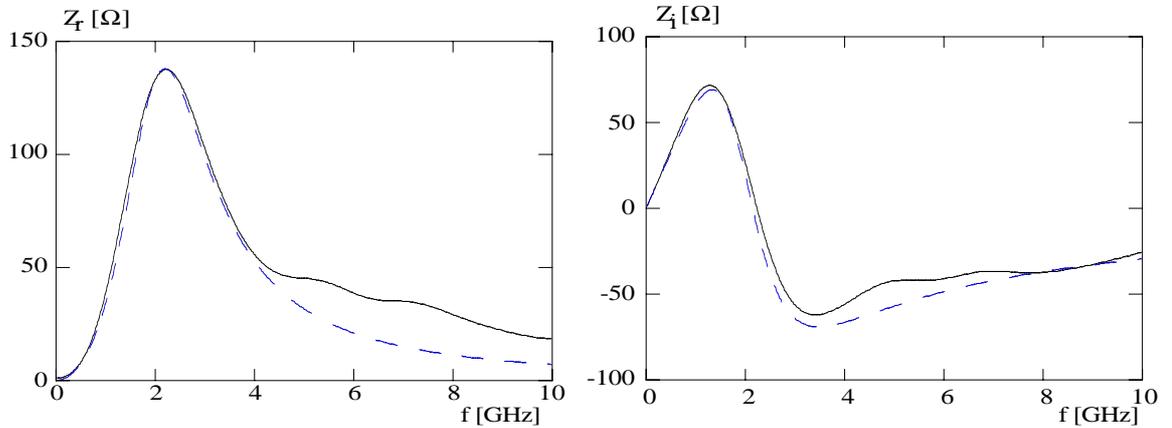

Fig. 30 - FFT of the wake function over 30 cm (solid). B. B. resonator impedance (dashed)

The parameters of the Broad-Band resonator can be found by measurements, for example, by measuring bunch lengthening in an accelerator. At the project stage these three parameters can be evaluated by comparing the loss factor dependence on the bunch length $K$ or wake potentials along the bunch $W_z(\tau)$, calculated numerically, with those corresponding to the broad-band resonator.

For the example in Fig. 30 we got Q = 1, R = 138Ω and $f_r$ = 2.2 GHz. However, the broad-band resonator model has some disadvantages which can become important for the extremely short or very long bunches.

Asymptotic frequency dependence of the broad-band resonator impedance is $\omega^{-2}$, while analytical study shows that the real part of the impedance decreases with frequency as $\omega^{-1/2}$ for a cavity with attached tubes and as $\omega^{-3/2}$ for an array of periodic cavities. This leads to incorrect energy loss estimate in the model, especially for very short bunches when the main contribution to the total energy loss is given by the high-frequency tail of the longitudinal impedance.

For very long bunches, according to the broad band resonator model, the loss factor decreases with the third power of the bunch length, while numerical calculations with the time domain codes shows that the loss factor of realistic vacuum chamber elements drops much faster, in an exponential fashion, with increasing σ for long bunches with Gaussian distribution.

## 8.2 Hofmann-Zotter impedance model

Two pairs of the broad band impedance models were proposed by A. Hofmann and B. Zotter [52] to overcome the limitations of the broad band resonator model in the high frequency and low frequency regions. Both can be adjusted to have either $\omega^{-1/2}$ or $\omega^{-3/2}$ asymptotic behaviour of the real part of the impedance at high frequencies. In the low frequency region, for the first pair of the impedance models the real part increases as $\omega^2$ and, for the special choice of the model parameters, as $\omega^4$ providing faster decrease of the loss factor with bunch length ($\sigma^{-5}$) than in the broad-band resonator model. The second improved impedance model yields even more faster decrease of the loss factor with bunch length, having an exponential character for very long bunches.

### 8.2.1 The first improved impedance model

The expression for the impedance model 1a) is:

$$Z(x) = R\left[\frac{1}{u}\sqrt{\frac{2}{u+1}} - \frac{\alpha^2}{x^2+\alpha^2}\right] + jR\left[\frac{1}{u}\sqrt{\frac{2}{u-1}} + \frac{\alpha^2}{x^2+\alpha^2} - \frac{2}{x}\right] \qquad (218)$$

where:

$$x = \frac{\omega}{\omega_1}, \quad u = \sqrt{x^2+1}, \quad \alpha = \frac{\omega_2}{\omega_1} \qquad (219)$$

The frequencies $\omega_1$, $\omega_2$ and the shunt impedance R are to be chosen to fit the impedance of a particular structure. For very high frequencies, the asymptotic behaviour of the impedance is given by:

$$Z_r(x) \sim x^{-3/2}, \quad Z_i(x) \sim \frac{2-\alpha}{x} \tag{220}$$

At low frequencies:

$$Z_r(x) \sim \left(\frac{1}{\alpha^2} - \frac{5}{8}\right)x^2, \quad Z_i(x) \sim \left(\frac{1}{\alpha} - \frac{3}{4}\right)x \tag{221}$$

For a special choice $\alpha = \sqrt{8/5}$ the real part of the impedance increases as $\omega^4$ providing a fast decrease of the loss factor with bunch length $\sim \sigma^{-5}$. In order to provide the high frequency asymptotic behaviour $\omega^{-1/2}$ a different model was proposed:

$$Z(x) = R\left[\frac{\sqrt{u^2+1}}{u\sqrt{2}} - \frac{\alpha^2}{x^2+\alpha^2}\right] - jR\left[\frac{\sqrt{u-1}}{u\sqrt{2}} - \frac{\alpha x}{x^2+\alpha^2}\right] \tag{222}$$

with:

$$x = \frac{\omega_1}{\omega}, \quad u = \sqrt{x^2+1}, \quad \alpha = \frac{\omega_2}{\omega_1} \tag{223}$$

At low frequencies the impedance has the following behaviour:

$$Z_r(x) \sim \left(\frac{1}{\alpha^2} - \frac{3}{8}\right)x^2, \quad Z_i(x) \sim \left(\frac{1}{\alpha} - \frac{1}{2}\right)x \tag{224}$$

8.2.2 The second improved impedance model

While the first impedance model describes well high frequency behaviour, it still can not supply an exponential fall-off of the loss factor with bunch length for long Gaussian bunches. In order to obtain this the real part of the impedance should vanish completely below a certain "cut-off" frequency (as it is for a realistic structure where the impedance is zero below the lowest mode). The second model impedances are chosen to satisfy this condition.

The 2a) real model impedance is given by the expression:

$$Z_r(x) = R\frac{\sqrt{|x|-1}}{x^2}; \quad |x| > 1$$

$$Z_r(x) = 0; \quad |x| < 1 \tag{225}$$

It has a maximum at x = ω/ω$_1$ = 4/3 and becomes proportional to ω$^{-3/2}$ at high frequencies. The Hilbert transform gives the impedance imaginary part:

$$Z_i(x) = \frac{R}{x^2}\left[\sqrt{1+x} - \sqrt{1-x} - x\right]; \quad |x| < 1$$

$$Z_i(x) = \frac{R}{x^2}\left[\sqrt{1+x} - x\right]; \quad |x| > 1$$

(226)

A similar impedance model (2b) has an asymptotic decrease of the real part proportional to the inverse square root of frequency at high frequencies. The real part is given by:

$$Z_r(x) = R\frac{\sqrt{|x|-1}}{|x|}; \quad |x| > 1$$

$$Z_r(x) = 0; \quad |x| < 1$$

(227)

which has a maximum at x=2. The expression for the imaginary part is:

$$Z_i(x) = \frac{R}{x}\left[2 - \sqrt{1+x} - \sqrt{1-x}\right]; \quad |x| < 1$$

$$Z_i(x) = \frac{R}{x}\left[2 - \sqrt{1+x}\right]; \quad |x| > 1$$

(228)

The loss factor for both models, 2a and 2b, can be easily obtained by numerical integration. It drops exponentially with bunch length for the two models for long Gaussian bunches.

### 8.3 Heifets-Bane impedance model

Recently, a new broad-band impedance model was proposed by S. Heifets [53] as the further development of K. Bane's approach used in his analysis of the impedance of the SLC damping ring [54,55]. The longitudinal impedance is described phenomenologically by expansion over √ω :

$$Z(\omega) = j\omega L + R + (1 + j\,sign(\omega))\sqrt{|\omega|}B + \frac{1 - j\,sign(\omega)}{\sqrt{|\omega|}}\tilde{Z}_c + \ldots \quad (229)$$

This model has been applied to estimate the broad-band impedance for SLAC B-factory and for the DAΦNE main rings (Φ-factory) [56]. Such an impedance model has two attractive features:
- First, the different terms of the expansion have a clear physical interpretation, describing correctly particular impedance-generating elements.
- Second, expressions for wake-fields and loss factors can be easily found analytically, simplifying the fitting procedure to extract the model parameters.

The first term of the expansion represents a low frequency inductive impedance. This impedance is typical for tapers, shielded bellows and vacuum ports, small discontinuities as slots, shallow cavities in flanges, shallow collimators and so on. Often, these elements give the main contribution to the impedance, leading to the excess bunch lengthening as in the case of SLC damping rings. For the small discontinuities the impedance remains inductive up to rather high frequencies.

The wake function for a Gaussian bunch corresponding to the *inductive* term is given by:

$$W_z(\tau) = -\frac{L\tau}{\sqrt{2\pi}\sigma_\tau^3} exp\left\{-\frac{1}{2}\left(\frac{\tau}{\sigma_\tau}\right)^2\right\} \qquad (230)$$

It has a minimum (maximum) at $\tau = -\sigma_\tau(+\sigma_\tau)$:

$$W_{max} = -W_{min} = \frac{L}{\sqrt{2\pi e}\sigma_\tau^2} \qquad (231)$$

The plot of $-W_z(\tau)/|W_z|_{max}$ which is suitable to compare with TBCI [57] or ABCI [25] code results for azimuthally symmetric structures or MAFIA [58] for 3D structures is shown in Fig. 31.

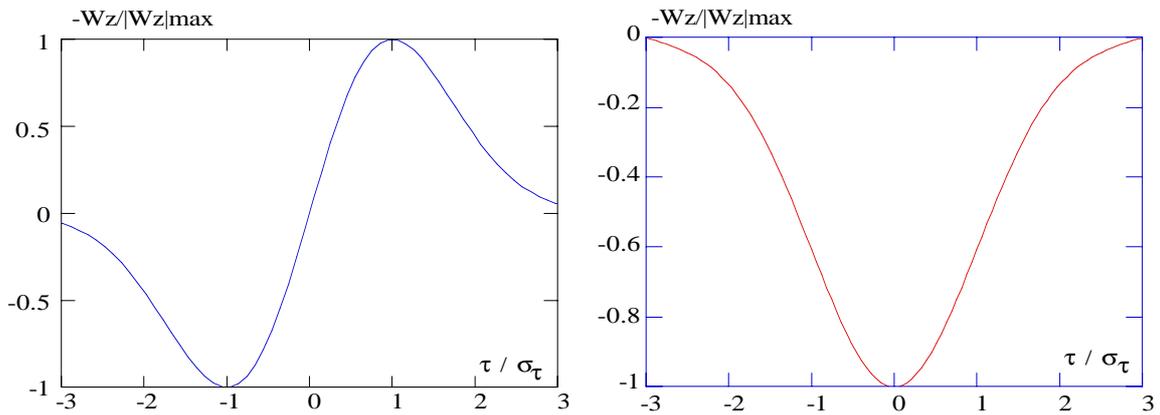

Fig. 31 - Normalized inductive wake functions: inductive and resistive

Eq. (231) allows to extract straightforward the inductance L from numerical results. The loss factor for the inductive impedance is zero. An example of a shallow cavity and corresponding wake are shown in Fig. 32.

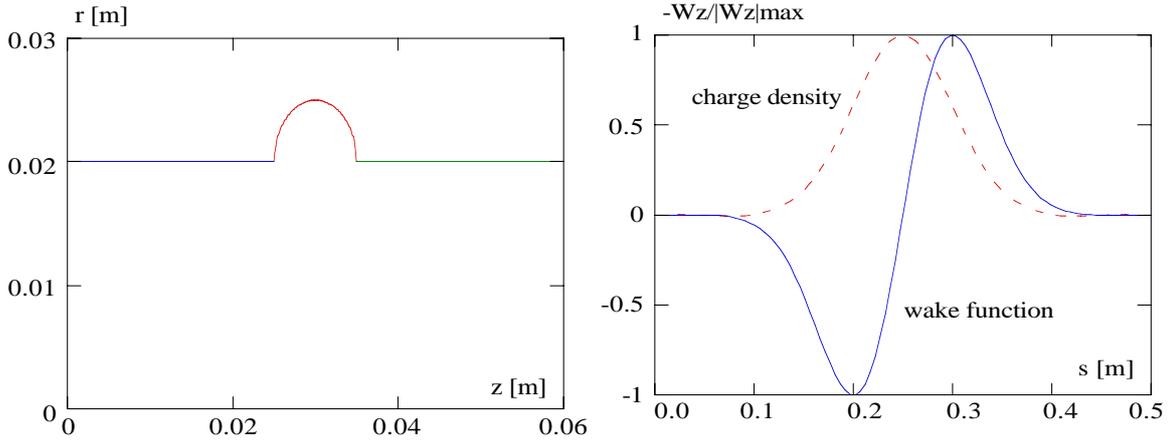

Fig. 32 - Shallow cavity and wake function for $c\sigma_\tau = 5$ cm

The wake-field for the *resistive* term $Z(\omega) = R$ is:

$$W_z(\tau) = -\frac{R}{\sqrt{2\pi}\sigma_\tau} exp\left\{-\frac{1}{2}\left(\frac{\tau}{\sigma_\tau}\right)^2\right\} \qquad (232)$$

The ratio $-W_z(\tau)/|W_z|_{max}$ is shown in Fig. 33. In storage rings such a wake is expected for deep cavities for bunches with length comparable with the beam pipe radius. Fig. 33 reproduces an example of wake for SLC damping ring cavity [55].

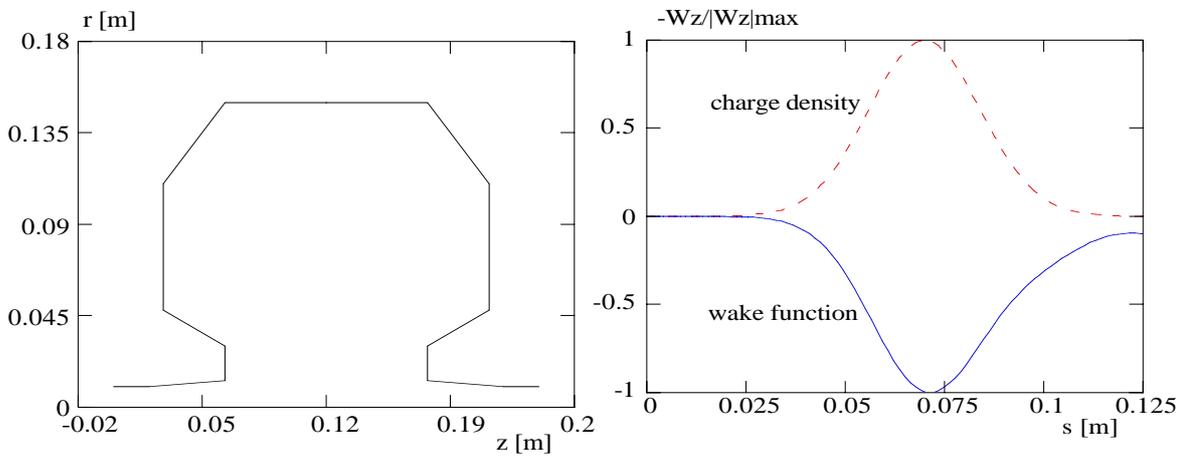

Fig. 33 - Deep cavity and wake function for $c\sigma_\tau = 1.4$ cm

For good resistors the expression for the loss factor has a simple form:

$$K(\sigma_\tau) = \frac{R}{2\sqrt{\pi}\sigma_\tau} \qquad (233)$$

The third term has the structure of the resistive wall impedance with the following wake-function (See Fig. 34 a):

$$W_z(\tau) = \frac{B|\tau|^{3/2}}{4\sigma_\tau^3}\{I_{-3/4}(\eta) - I_{1/4}(\eta) - I_{-1/4}(\eta) \pm I_{3/4}(\eta)\}e^{-\eta} \qquad (234)$$

where $\eta = \left(\frac{\tau}{2\sigma_\tau}\right)^2$ and the $\pm$ sign corresponds to positive and negative "$\tau$", respectively. $I_\nu$ are the modified Bessel functions of fractional order. Applying the loss factor definition we have:

$$K(\sigma_\tau) = \frac{B}{\pi\sigma_\tau^{3/2}}\frac{\Gamma(3/4)}{2} \qquad (235)$$

where $\Gamma(3/4)/2 = 0.6127...$.

The forth term in (229) has the same dependence on $\omega$ as the impedance of a cavity with attached tubes at high frequencies:

$$Z(\omega) = \frac{(1 - jsign(\omega))}{\sqrt{|\omega|}}\tilde{Z}_c \qquad (236)$$

The wake-function corresponding to the impedance is given by:

$$W_z(\tau) = \frac{\tilde{Z}_c}{2\sqrt{\sigma_\tau}}\sqrt{\frac{|\tau|}{\sigma_\tau}}\{I_{-1/4}(\eta) \pm I_{1/4}(\eta)\}e^{-\eta} \qquad (237)$$

$\pm$sign stands for positive and negative "$\tau$". The function $-W_z(\tau)/|W_z|_{max}$ is shown in Fig. 34b.

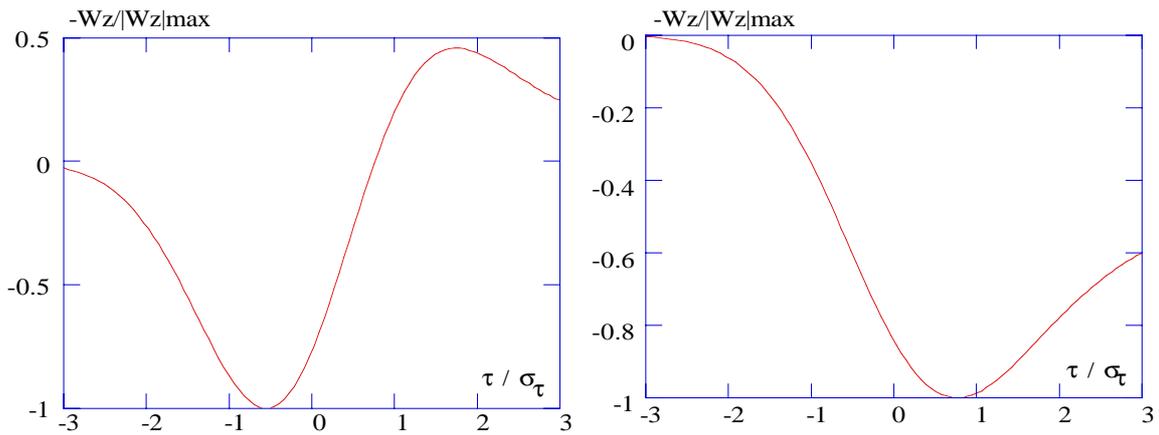

Fig. 34 - Normalized wake functions corresponding to the cases:

a) $(1 + jsign(\omega))\sqrt{|\omega|}B$    b) $\frac{1 - jsign(\omega)}{\sqrt{|\omega|}}\tilde{Z}_c$

Using (26) we obtain the expression for the loss factor:

$$K(\sigma_\tau) = \frac{B}{\pi \sigma_\tau^{3/2}} \frac{\Gamma(3/4)}{2} \qquad (238)$$

with $\Gamma(1/4)/2 = 1.8128...$

The parameters L, R, B, Zc of the broad-band model are extracted from TBCI, ABCI or MAFIA results by fitting the numerical functions $W(\tau)$ and $K(\sigma_\tau)$ to above analytical expressions. In Fig. 35 we show an example of wake function for a cavity with attached tubes where the high frequency behaviour is dominant.

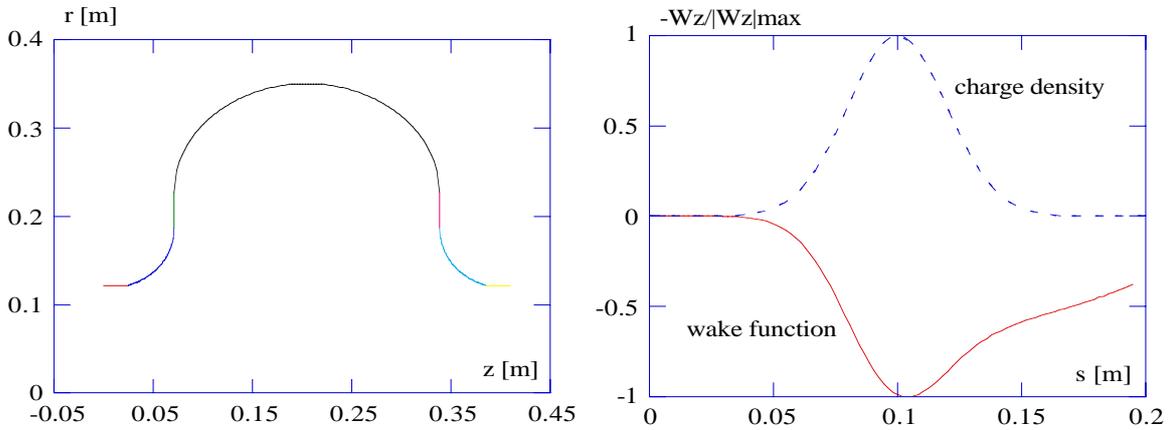

Fig. 35 - Cavity with attached tubes and wake function for $c\sigma_\tau = 2$ cm

## Acknowledgments

We wish to thank M. Serio for some enlightening comments and suggestions.

# - 68 -

# INDEX